\documentclass[twocolumn]{aastex631}

\usepackage{amsmath}
\usepackage{comment}

\newcommand{\pfrac}[2]{\left( \frac{#1}{#2} \right)}
\newcommand{\pdiff}[2]{\frac{\partial #1}{\partial #2}}

\newcommand{\rp}{r_{\rm p}}

\newcommand{\rtidal}{r_{\rm t}}

\newcommand{\tgw}{\tau_{\rm GW}}

\newcommand{\MBH}{M_{\bullet}}

\newcommand{\Sigmad}{\Sigma_{\rm d}}

\begin{document}

\title{Delayed Radio Flares in Tidal Disruption Events from Star-Disk Collision Outflows}

\correspondingauthor{Itai Linial}
\email{il2688@nyu.edu, il2432@columbia.edu}

\author[0000-0002-8304-1988]{Itai Linial}\thanks{NHFP Einstein Fellow}
\affiliation{Department of Physics and Columbia Astrophysics Laboratory, Columbia University, New York, NY 10027, USA}
\affiliation{Center for Cosmology and Particle Physics, Physics Department, New York University, New York, NY 10003, USA}

\author[0000-0002-4670-7509]{Brian D.~Metzger}
\affil{Department of Physics and Columbia Astrophysics Laboratory, Columbia University, New York, NY 10027, USA}
\affil{Center for Computational Astrophysics, Flatiron Institute, 162 5th Ave, New York, NY 10010, USA} 

\author[0000-0001-5660-3175]{Andrei M.~Beloborodov}
\affil{Department of Physics and Columbia Astrophysics Laboratory, Columbia University, New York, NY 10027, USA}
\affiliation{Max Planck Institute for Astrophysics, Karl-Schwarzschild-Str. 1, D-85741, Garching, Germany}

\begin{abstract}
A growing fraction of tidal disruption events (TDEs) exhibit radio emission that rises only years after the optical or infrared flare, indicating delayed outflow activity. In some events the outflow is inferred to be slow ($\sim 0.02 \, c$) and massive ($\gtrsim 0.01-0.1 M_{\odot}$), challenging models such as delayed jets and disk state transitions. We propose a new mechanism for such delayed outflows: repeated collisions between a TDE accretion disk and a pre-existing stellar extreme-mass-ratio-inspiral (EMRI) orbiting the black hole. In this scenario, the delay reflects the viscous time required for the initially compact TDE disk to expand and intercept the EMRI orbit, rather than delayed jet launching or off-axis viewing effects. Once star-disk collisions commence, repeated impacts eject outflows with velocities comparable to the orbital speed, $v_{\rm w} \sim 0.02-0.1c$. We develop a time-dependent model for the coupled evolution of the spreading disk and EMRI-induced mass-loss, identifying regimes where the outflow is dominated by disk material or ablated stellar debris. Depending on disk viscosity, orbital period, and collision efficiency, masses $\sim (10^{-3}-1) \, \rm M_\odot$ can be launched with energies up to $10^{51} \rm \, erg$, years after the TDE. These outflows produce radio emission through interaction with circumnuclear material or earlier TDE ejecta, consistent with observed late-time radio re-brightening. This model predicts a connection between delayed radio flares and EMRI-hosting systems, potentially including those exhibiting quasi-periodic eruptions (QPEs) powered by star-disk collisions, though the conditions for bright radio flares may not always match those necessary for detectable QPEs.
\end{abstract}

\keywords{Supermassive black holes (1663), Tidal disruption (1696), Radio transient sources (2008)}

\section{Introduction} \label{sec:intro}

Radio emission from Tidal Disruption Events (TDEs) probes the outflows from these systems as they interact with the circumnuclear medium (CNM) or other gas in the vicinity of the supermassive black hole (SMBH; \citealt{Giannios&Metzger11,Metzger+12,Berger+12}). A small fraction ($\lesssim 1\%$) of TDEs generate prompt and powerful collimated relativistic jets, typically viewed as hard X-ray sources on-axis or nearly on-axis \citep{Bloom+11,Burrows+11,Levan+11,Cenko12,Brown15,Kawamuro16,Andreoni+22,Pasham+23}, that also power luminous synchrotron radio emission starting almost immediately after the disruption \citep{Zauderer+11,Berger+12,Mimica+15}.  

However, in addition to these conspicuously jetted events, a growing sample of TDEs show radio emission consistent with non- or mildly-relativistic outflows (e.g., \citealt{Alexander16, vanVelzen16,Alexander+17,Saxton17,Mattila18}; see \citealt{Alexander+20} for a review), whose light curves in many cases only begin to rise following a significant delay after the optical peak (\citealt{Horesh+21,Horesh+21b,Perlman+22,Cendes+22,Sfaradi+22}). \citet{Cendes+24} found that approximately $40\%$ of optically-selected TDEs exhibit late-time radio flares, delayed by hundreds to thousands of days after the disruption (see also \citealt{Alexander+26,Goodwin+25a}). Complicating the phenomenology further, several TDEs show multiple distinct peaks in their radio light curves on different timescales, e.g., ASASSN-15oi \citep{Horesh+21,Hajela+25}, AT 2019azh \citep{Sfaradi+22}, AT 2020vwl \citep{Goodwin+23}, 
and the off-nuclear event AT 2024tvd \citep{Sfaradi+25}.

Several explanations have been put forward for delayed radio emission in TDEs.  The afterglow from a relativistic jet could rise late, either because the jet is viewed off-axis \citep{Giannios&Metzger11,Beniamini+23,Matsumoto&Piran24,Sfaradi+24}, or because the jet launching is intrinsically delayed after the disruption.  Delayed jet launching could occur if the accretion rate onto the SMBH peaks only well after the optical flare \citep{Gezari+17,Metzger22,Alexander+26}; if the jet is activated by a late-time state transition in the disk \citep{Giannios&Metzger11,GoodwinMummery26}; or if the jet is delayed in escaping from the dense gaseous environment encasing the black hole \citep{Tchekhovskoy+14,Teboul&Metzger23,Lu+24}. 
Jets or outflows launched close to the black hole should possess high outflow speeds, $v_{\rm w} \gtrsim 0.2 \, c$ (e.g., \citealt{Jiang+19}), with relatively low ejecta masses, capped by that of the inner TDE disk.

\begin{figure*}
    \centering
    \includegraphics[width=\linewidth]{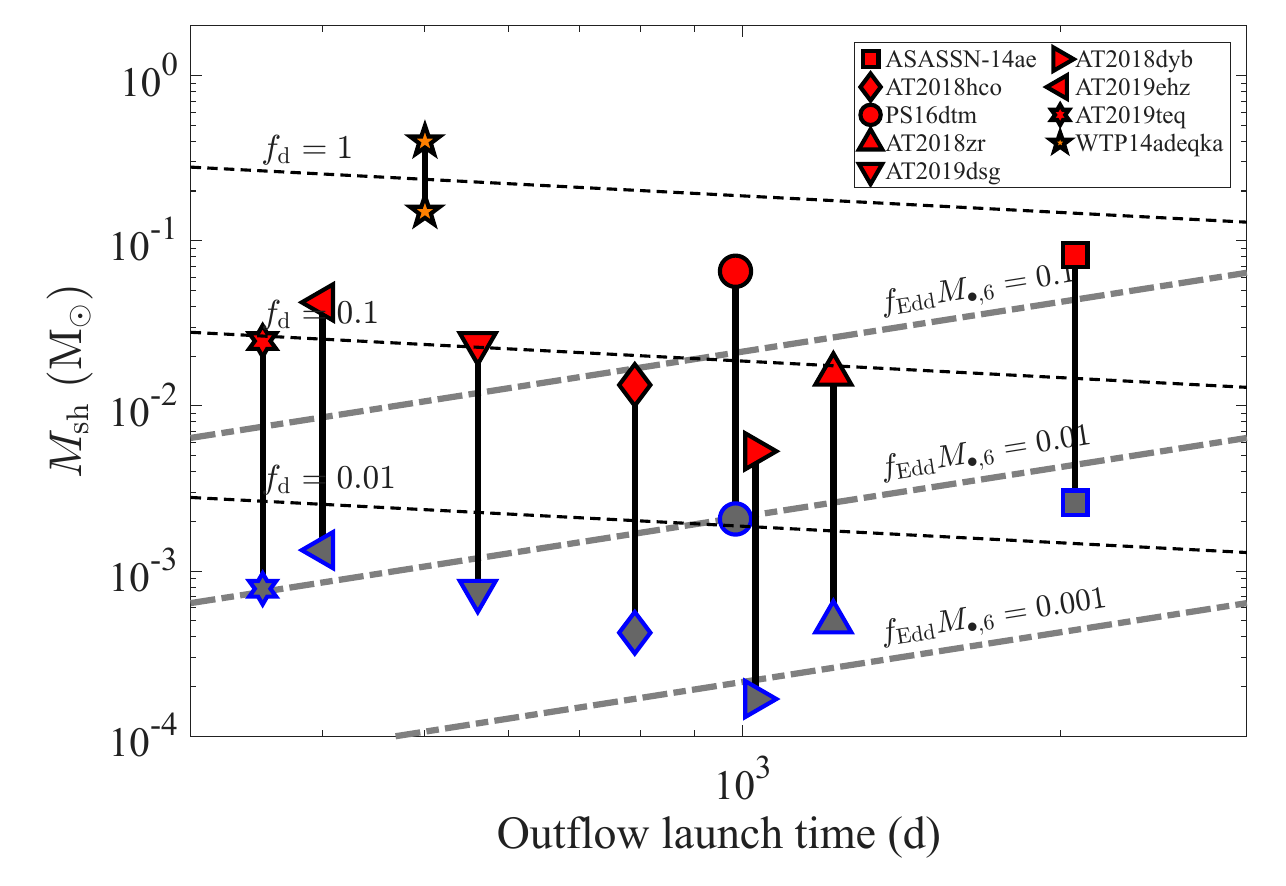}
    \caption{Shocked mass versus launch times of 9 TDEs exhibiting delayed radio flares (8 from \citealt{Cendes+24} and the remaining one from \citealt{Golay+25}). Shocked masses $M_{\rm sh}=2E_{\rm k}/v_{\rm w}^2$ were derived from the equipartition analysis of \cite{Cendes+24}. Blue and red markers correspond to a range of assumptions on the microphysical equipartition parameters, with $\epsilon_{\rm e}=\epsilon_{\rm B}=0.1$ (blue markers), and $\epsilon_{\rm e}=10^{-2}$, $\epsilon_{\rm B}=10^{-3}$ (motivated by first principle simulations of magnetic field amplification and electron acceleration in non-relativistic shocks, e.g., \citealt{Caprioli+14}; red markers), scaling as $M_{\rm ej} \propto \epsilon_{\rm e}^{-0.5} \epsilon_{\rm B}^{-8/17}$ \citep[e.g.,][]{BarniolDuran&Piran13,Matsumoto&Piran21}. For WTP14adeqka, VLBA observations and the measurement of the synchrotron cooling break frequency lift some of the degeneracies, and the range of values correspond to the uncertainty in expansion velocity. The total remaining TDE disk mass as a function of time is plotted in black dashed lines, taking an initial mass $m_{\rm d,0}=f_{\rm d}({\rm M_\odot}/2)$, initial spreading time $t_{\rm v,0}=50 \, \rm d$ and $m_{\rm d} \propto t^{-1/3}$. Dashed-dotted gray lines show the total disk mass at time $t$, assuming that $\dot{m}_{\rm d}(t)= f_{\rm Edd} \dot{M}_{\rm Edd}$ where $\dot{M}_{\rm Edd}$ is the Eddington accretion rate of an SMBH of mass $10^6 \, \rm M_\odot$.}
    \label{fig:OutflowMassRadioTDEs}
\end{figure*}

Yet, in many cases, the inferred outflow velocities 
are much slower, $v_{\rm w} \approx 0.02-0.1 \, c$, and carry substantial mass (\citealt{Cendes+24}, and Fig.~\ref{fig:OutflowMassRadioTDEs}). A recent well-studied example is the the dust-obscured (infrared-discovered) TDE WTP14adeqka \citep{Golay+25}, whose radio emission started rising $\sim 4$ years after the infrared flare detection, before peaking at $\sim$ 6.5 years and then declining thereafter. Very Long Baseline Array (VLBA) imaging at multiple epochs were taken roughly 9 years after disruption, resolving the source size and measuring expansion. \citet{Golay+25} inferred a non‐relativistic outflow of velocity $0.02-0.04c$, equipartition energy $\sim 10^{51}$ erg, carrying a mass $\gtrsim 0.1M_{\odot}$. The small spatial size and lack of astrometric shift argue against a promptly launched, off-axis relativistic jet scenario (which would predict larger size) for this event, but instead support a slow outflow launched with an intrinsic delay of $\sim 2$ years after the disruption.

Several sources of slower outflows can arise in TDEs from matter ejected promptly during the disruption \citep{Krolik+16,Yalinewich+19,Lei+24,Hu+25}, in winds launched during the disk formation process \citep{Metzger22,Price+24,Tuna25}, or in thermal state transitions \citep{Wu_2026}. However, it is unclear why these slow-outflows should turn on suddenly several years after the disruption and carry so much mass, sometimes approaching that of the entire TDE disk. Fig.~\ref{fig:OutflowMassRadioTDEs} shows the swept up mass, as derived from the radio equipartition analysis presented in \cite{Cendes+24,Golay+25}. Dashed lines show the total available mass budget in the disk at the time the outflow was launched -- dashed black lines show the declining TDE disk mass of a $1 \, {\rm M_\odot}$ star, assuming that the disk formed with a fraction $f_{\rm d}$ of the bound TDE debris. Dash-dotted lines show the necessary disk mass, assuming that at time $t$ the accretion rate is a fraction $f_{\rm Edd}$ of the Eddington accretion rate of a $10^6 \, {\rm M_\odot}$ SMBH. At least in some cases, the inferred mass is seen to be a substantial fraction of the total mass budget of the disk, or even exceed it (e.g., for $f_{\rm d}\lesssim0.1$ or $f_{\rm Edd} M_{\bullet,6}\lesssim 0.01-0.1$). If these sources have not experienced substantial deceleration yet (as argued by \citealt{Cendes+24}), the swept up mass inferred from equipartition analysis represents a lower limit on the initial ejecta mass in the delayed outflow, thereby posing a challenge to models that invoke the TDE accretion disk as the ultimate mass reservoir of outflows, especially since only a small inner region of the disk is expected to be ejected during state transitions \citep[e.g.,][]{Done+07}.

A new class of nuclear transients known as quasi-periodic eruptions (QPEs) has recently been identified through their repeating, soft X-ray flares with recurrence times of hours to days (e.g., \citealt{Miniutti+19,Giustini_2019,Arcodia+21,Chakraborty_2021}). QPEs are observed in otherwise quiescent galactic nuclei hosting low-mass SMBH, and are characterized by large-amplitude X-ray variability with relatively stable periods over many years \citep[e.g.,][]{Arcodia+24b}. While several theoretical interpretations for QPEs have been proposed, a promising framework links them to repeated interactions between a compact object or star on a tight orbit (an extreme mass-ratio inspiral; EMRI) and a gaseous disk surrounding the central SMBH (e.g., \citealt{Xian_2021}). In this picture, collisions between an orbiting star and a disk can generate recurrent shocks and debris streams, naturally producing quasi-periodic X-ray flares synchronized with the orbital period (\citealt{Linial_Metzger_23,Tagawa&Haiman23,Franchini_23,Linial_25,Huang+25}; although, see \citealt{King_2020,Pan+22,Middleton+25}). If the disk is formed by the tidal disruption of a separate star, this scenario directly connects QPEs to TDEs. Strong observational support for this connection has recently emerged with the detection of QPEs in the late-time aftermath of TDE flares (e.g., \citealt{Chakraborty_2021,Quintin+23,Nicholl+24,Chakraborty+25,Bykov+25}).

Motivated by the emerging connection between QPEs, TDE disks, and star-disk interactions, we show here that repeated collisions of a stellar EMRI with a spreading TDE accretion disk can naturally power delayed outflows and late radio flares. The delay is set by the viscous spreading time of the initially compact TDE disk to expand outward and intercept the EMRI orbit, rather than delayed jet launching or viewing-angle effects. Once collisions commence, disk material and ablated stellar debris can be ejected at velocities comparable to the EMRI orbital speed $v_{\rm w} \lesssim 0.1 c$, producing a slow but energetic outflow capable of powering radio emission as it interacts with the CNM or earlier TDE ejecta. Our framework introduces the EMRI as an additional mass reservoir for launching delayed outflows, thereby avoiding the limited mass available in the accretion disk, which is potentially in tension with that required in some sources (e.g., Fig.~\ref{fig:OutflowMassRadioTDEs}).

This paper is organized as follows. Sec.~\ref{sec:conceptual} conceptually overviews the star--disk interaction scenario and its observational signatures. Sec.~\ref{sec:model} describes the key components of the model, including the spreading TDE disk, the stellar EMRI, and mass ejection during repeated collisions. In Sec.~\ref{sec:diskevo}, we present a time-dependent model for the coupled evolution of the disk and EMRI and derive the properties of the resulting outflow. In Sec.~\ref{sec:radio}, we compute the radio emission produced as the delayed outflow interacts with its environment and compare our predictions with observed late-time radio flares in TDEs. In Sec.~\ref{sec:implications}, we discuss the implications of this mechanism for TDE and QPE demographics, connections between X-ray and radio variability, and prospects for future observations. We summarize our conclusions in Sec.~\ref{sec:conclusions}. Glossary of used symbols is provided in Appendix \ref{sec:Glossary}.

\section{Conceptual Overview}
\label{sec:conceptual}

Dynamical processes in the dense stellar environments of galactic nuclei occasionally bring stars close to the central SMBH (e.g., \citealt{Alexander_2005,Sari_Fragione_2019}). Angular momentum relaxation drives stars from a parsec-scale to nearly radial trajectories that approach the SMBH to within the tidal radius $r_{\rm t} \sim \rm au$, resulting in a TDE \citep{Rees_1988}, at a typical rate $\mathcal{R}_{\rm tde} \approx 10^{-5}-10^{-4} \, \rm yr^{-1}$ per galaxy \citep[e.g.,][]{Magorrian&Tremaine99,Stone_2020}. A different evolutionary fate awaits stars on initially tighter orbits: rapid dissipation of orbital energy through gravitational wave (GW) emission or hydrodynamical drag may efficiently circularize orbits before they are tidally disrupted, bringing them towards the SMBH as low-eccentricity, extreme-mass ratio inspirals (EMRIs; \citealt{Linial_Sari_2017, Sari_Fragione_2019,Linial_Sari_2023}). 
One way to feed stars onto such EMRI-producing trajectories is via the \citet{Hills_1988} mechanism, in which a stellar binary scattered towards the SMBH is tidally dissociated, unbinding one member and leaving the other star on a more tightly bound orbit.

Interestingly, TDEs and EMRIs may temporally coincide and interact: the long GW inspiral timescale of stellar-EMRIs at radii $a_0 \gtrsim \rtidal$ and 
orbital periods of hours-days ($\tgw\gtrsim 10^6-10^7 \, \rm yr$, e.g., \citealt{Linial_Quataert_24b}) is much longer than $\mathcal{R}_{\rm tde}^{-1}$, the mean time between TDEs, implying that during its slow inspiral, a single EMRI will likely witness many TDEs while its semi-major axis is comparable to $\rtidal$ (\citealt{Linial_Metzger_23}).

\begin{figure*}
    \centering
    \includegraphics[width=\linewidth]{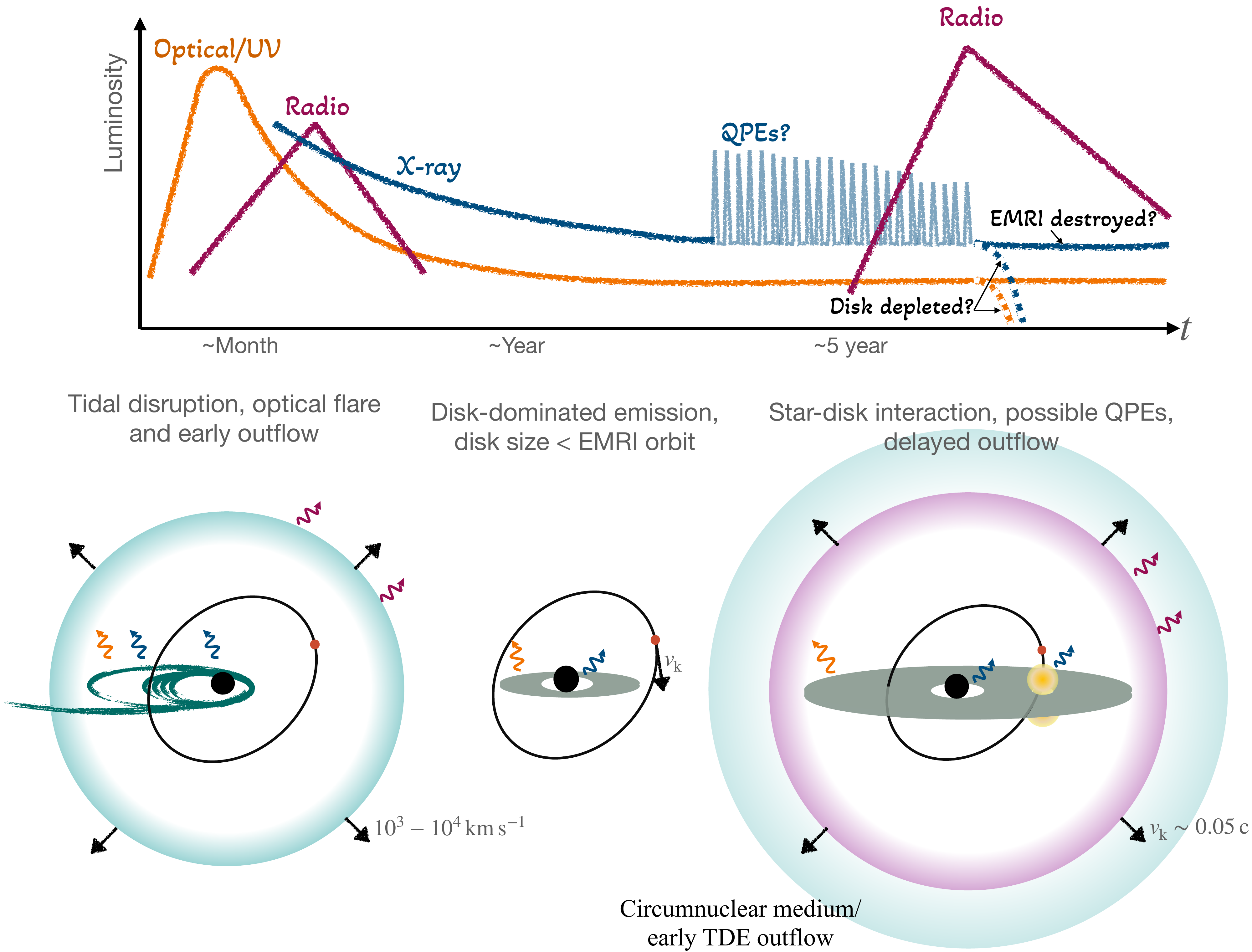}
    \caption{Key phases of the model, matched along the top by schematic lightcurves at different bands.  The early optical/UV flare, decaying X-ray emission and prompt early radio flare on timescales $\lesssim \rm yr$ are intrinsic to the TDE.  Once the disk spreads to meet the orbit of a stellar EMRI, additional emission signatures may occur over longer timescales $\gtrsim $ yrs from repeated disk-star collisions, including X-ray QPEs and delayed radio outflows.  Termination of the disk or QPE emission can arise if either the disk is depleted, or the EMRI destroyed, by the repeated collisions (Sec.~\ref{sec:Outflow_abl_domninated}).  }
    \label{fig:DelayedRadioSchematic}
\end{figure*}

Fig.~\ref{fig:DelayedRadioSchematic} schematically illustrates the key elements of our model and some of the predicted observable signatures of a TDE in the presence of a nearby EMRI:
\begin{itemize}
    \item \textit{Initial TDE flare}. Stellar tidal disruption produces a complex multi-wavelength electromagnetic transient evolving over a wide range of timescales, spanning months to decades. During their first year, TDEs typically exhibit a bright optical/UV flare, rapidly rising over a month, followed by a gradual decay, often accompanied by soft X-rays that subside over a $\sim$months decay timescale
    similar to that of the optical flare (e.g., \citealt{Gezari_2021}; though in many events the X-ray peak is substantially delayed; e.g., \citealt{Gezari+17,Guolo+24,Hajela+25}). The physical origin of early optical/UV and X-ray emission in TDEs remains elusive, largely due to the poorly understood role of e.g., debris self-intersection and circularization, super-Eddington accretion and radiation reprocessing. About $20\%$ of optically-selected TDE show a detectable radio flare during the first few months following the disruption of the star, indicative of a non-relativistic outflow of velocity $\lesssim {\rm few\times}10^4 \, \rm km\,s^{-1}$, launched promptly at the time of the early optical flare (e.g., \citealt{Alexander+26}). As this outflow shocks the surrounding medium, it powers synchrotron afterglow emission, detected at GHz frequencies. 
    
    During these early stages, any EMRI present around the SMBH will be largely unaffected, and the TDE would evolve as in the EMRI-free case (leftmost phase of Fig.~\ref{fig:DelayedRadioSchematic}).

    \item \textit{Disk-dominated phase}. Regardless of the details of the early gas and radiation dynamics, a significant fraction of the bound stellar debris is expected to eventually circularize and form a compact accretion disk feeding the SMBH, with an initial radial extent comparable to $\rtidal$, as illustrated in the center phase of Fig.~\ref{fig:DelayedRadioSchematic}. TDE emission at this late phase is well-described by a viscously spreading, radiatively cooled thin-disk model \citep[e.g.,][]{Mummery&VanVelzen24}. This phase often begins about a year after the initial TDE flare, when the optical/UV decline flattens into a plateau, while the soft X-ray emission continues to gradually decay. 

    \item \textit{Star-disk interaction and delayed outflows}. Initially, the EMRI is mostly unaffected by the TDE and its nascent compact accretion disk. However, as the disk feeds the SMBH, conservation of angular momentum requires the outer radius of the disk to viscously spread (e.g., \citealt{Cannizzo+90,Shen&Matzner14,WinterGranic&Quataert25}), until it eventually intercepts the EMRI orbit. Star-disk collisions ensue, ejecting disk material \citep{Huang+25} and ablating the star \citep{Yao_25}, and possibly powering recurrent QPE flares, seen in soft X-rays (\citealt{Linial_Metzger_23,Linial_25,Vurm+25}; rightmost phase of Fig.~\ref{fig:DelayedRadioSchematic}). While the ultimate fate of these ejecta depends on the details of the star-disk collisions (e.g., relative inclination, disk surface density and scale height), 
    a significant fraction is generally expected to become unbound, forming a delayed powerful outflow. The outflow typically carries a mass $m_{\rm w}\sim(10^{-3}-1) \, M_\odot$, with a velocity $v_{\rm w} \approx 0.02-0.1 \,  c$, corresponding to a kinetic energy $E_{\rm k} \gtrsim 10^{48}-10^{51} \, \rm erg$ (see Sec.~\ref{sec:diskevo} for details). 
    
    Depending on the depletion rate of the disk mass by stellar collisions, and the EMRI ablation rate, repeated collisions could destroy the star or the disk altogether, or both could survive over long timescales. These outcomes depend to leading order on the disk mass and its viscous spreading time, as well as the mass and radius of the stellar EMRI. If the disk is significantly depleted, its optical/UV/X-ray emission could be suppressed at late times (illustrated schematically with dashed lines in Fig.~\ref{fig:DelayedRadioSchematic}). The disk emission could also be attenuated if the collision-induced outflow shrouds the disk behind a high optical depth.

    The material ejected from the star-disk collisions expands and shocks the surrounding ambient medium, comprised of pre-existing CNM and earlier outflows from the TDE. Similar to the early radio emission from prompt outflows, the delayed EMRI-driven outflows power a distinct radio flare, often delayed by years after the TDE (Sec.~\ref{sec:radio}).
\end{itemize}

\section{Model components}
\label{sec:model}

\subsection{Tidal disruption and spreading accretion disk}

A main-sequence star of mass $m_\star^{\rm tde}$ and radius $R_\star^{\rm tde}$ undergoes tidal disruption as it approaches the SMBH of mass $\MBH$ on a parabolic orbit with pericenter distance $\rp \lesssim \rtidal \equiv R_\star^{\rm tde} (\MBH/m_\star^{\rm tde})^{1/3}$. Roughly half of the stellar debris are unbound, while the other half remains bound and falls back towards the SMBH on highly eccentric orbits. Subsequent energy dissipation (driven by, e.g., debris-stream self-intersection, \citealt{Bonnerot+17}; nozzle shocks, \citealt{Kochanek94}; or stream-disk collisions, \citealt{Steinberg&Stone24}) circularizes the bound debris, ultimately forming an accretion disk of mass $m_{\rm d,0} =f_{\rm d} m_\star^{\rm tde}/2$ (where $f_{\rm d} \leq 1$) and characteristic initial radius $r_{0} = 2\rp \lesssim 2 \rtidal$, set by conservation of angular momentum. As the disk feeds mass onto the SMBH, it viscously spreads outwards (e.g., \citealt{Cannizzo+90}), conserving its total angular momentum, $J_{\rm d} \propto m_{\rm d} r_{\rm d}^{1/2} = \rm const$. By the time the disk's outer edge reaches a distance $r_{\rm d} = a_0 > r_{0}$, its remaining mass will have decreased to
\begin{equation} \label{eq:m_disk_at_a0}
    m_{\rm d}(r_{\rm d}=a_0) = 2^{-1/2}f_{\rm d} m_\star^{\rm tde} (a_0/\rtidal)^{-1/2} \,.
\end{equation}
This represents an upper limit on the remaining disk mass reaching a given radius, as it neglects angular momentum lost from the system, for example, to magnetized disk winds \citep[e.g.,][]{Metzger+08}. We return to the accretion disk evolution in Sec.~\ref{sec:diskmodel}.

\subsection{Stellar EMRI}

Well prior to the TDE, a distinct star of mass $m_\star$ and radius $R_\star$ was brought in to the vicinity of the SMBH as an EMRI.  At the time of the TDE the EMRI resides on a nearly circular orbit of period $P_{\rm orb}$, radius $a_0$ and Keplerian velocity
\begin{equation} \label{eq:v_k_from_P}
    v_{\rm k} = \pfrac{2\pi G\MBH}{P_{\rm orb}}^{1/3} \approx 0.07\,c \; \pfrac{ M_{\bullet,6}}{P_{\rm orb,24}}^{1/3}  \,.
\end{equation}
where $P_{\rm orb} = {1 \, \rm d} \, P_{\rm orb,24}$.  Moderate eccentricities $e \sim 0.1$ are expected on radial scales corresponding to orbital periods of around a day for EMRIs supplied to the nucleus via the Hills mechanism and dynamical two-body scattering (e.g., \citealt{Linial_Sari_2022,Linial_Sari_2023}).

In vacuum, the EMRI's orbit decays through gravitational wave emission on a timescale
\begin{multline} \label{eq:tgw}
    \tgw \approx \frac{5}{64} \frac{G\MBH}{c^3} \pfrac{a_0}{R_{\rm g}}^4\pfrac{\MBH}{m_\star} \approx
    \\
    2\times 10^7 \, {\rm yr} \; m_1^{-1} M_{\bullet,6}^{-2/3} P_{\rm orb,24}^{8/3} \,,
\end{multline}
much longer than the interval $\sim 10^{4-5}$ yr between TDEs in a typical Galactic nucleus (e.g., \citealt{Magorrian&Tremaine99}). Many, if not most, TDEs are therefore expected to occur with an EMRI on a short period orbit present in the same Galactic nucleus \citep{Linial_Metzger_23}. Generally, the orbital plane of the EMRI will be inclined with respect to that of the disrupted star defining the TDE disk, as the two stars are brought towards the SMBH independently. 

\subsection{Mass Ejection from Star-Disk Collisions}
\label{sec:massejection}

Repeated collisions between the star and the disk ensue as the disk's extent approaches the EMRI orbit at radius $a_0$. With each collision, the EMRI excavates a small amount of disk mass
\begin{equation} \label{eq:delta_m_disk}
    \delta m_{\rm d} \approx \pi R_{\rm eff}^2 \Sigmad(a_0) \,,
\end{equation}
where $\Sigmad(a_0)$ is the disk's local surface density, and $R_{\rm eff}\gtrsim R_\star$ is the effective radius of the ``hole'' punctured by the star.

In addition to the excavated disk mass, star-disk collisions act to \textit{ablate} the EMRI, as a result of the reverse shock driven into its outer envelope.  Based on hydrodynamic simulations of star-disk collisions, \citet{Yao_25} found that after several collisions the stripped mass per collision approaches a steady value:
\begin{multline} \label{eq:delta_m_abl}
    \delta m_{\star} \approx \eta \frac{\pi R_\star^4 (\Sigmad/H_{\rm d}) v_{\rm k}^2}{Gm_\star} \approx \\
    \eta \,\pi R_\star^2 \Sigmad(a_0) \frac{\rtidal}{a_0} \pfrac{\MBH}{m_\star}^{2/3} \,,
\end{multline}
where $\rtidal$ is the tidal radius of the EMRI, and $\eta=0.03$ is the dimensionless stripping efficiency for an $n = 3/2$ polytrope, as appropriate for low-mass (fully convective) stars \citep{Yao_25}. The estimate (Eq. \ref{eq:delta_m_abl}) assumes the disk scale height, $H_{\rm d}$, is comparable to the star's radius (\citealt{Yao_25}, their Eq.~5).  The parameter scalings follow from assuming that a fixed fraction of the shocked stellar mass, estimated by balancing the ram pressure to the local hydrostatic pressure below the stellar surface, becomes unbound with each collision.   

The ablated material propagates from the stellar surface ballistically and quasi-spherically, until reaching the EMRI's Hill sphere, where its subsequent evolution is governed by the SMBH's tidal gravity, forming elongated streams of stellar debris \citep{Yao_25,Linial_25}. These streams then collide with the accretion disk, half an orbit later, possibly producing a bright flare that briefly outshines the disk in the soft X-rays \citep[a model for QPEs, see][]{Linial_Metzger_23,Linial_25}. Consequently, the EMRI's Hill sphere may accumulate considerable amount of slowly moving ablated material, thus increasing the effective radius for the disk interaction to $R_{\rm eff} \approx r_{\rm H}\equiv a_0 (m_\star/\MBH)^{1/3} = R_\star (a_0/\rtidal)$. However, at sufficiently long orbital periods, $P_{\rm orb}\gg 1.3 \, \rm d$, the ablated material within the Hill sphere becomes too dilute to eject the disk mass contained in the impact footprint, and the effective cross section radius reverts back to the star's physical size, i.e., $R_{\rm eff} \approx R_\star$ (\citealt{Linial_25}; their Eq.~37). In what follows we assume $R_{\rm eff}=R_\star$, but note this may underestimate the amount of impacted disk material at short orbital periods.

Both mass components ($\delta m_{\rm d}$, $\delta m_\star$) involve collisions with the rotating disk at roughly the local Keplerian velocity, $v_{\rm k}$ (for a typical value of the relative inclination of order unity). While a detailed study of the collision hydrodynamics is beyond the scope of this work, we anticipate that as long as the star's orbital plane is not closely aligned with the disk, a fraction of the shocked (star or disk) material is accelerated to beyond the local escape speed, $v_{\rm ej} \gtrsim v_{\rm k}\sqrt{2}$ \citep[as confirmed in simulations, e.g.,][]{Yao_25,Huang+25}. Many individual ejections, occurring every $\sim P_{\rm orb}/2$ accumulate to produce a coherent unbound outflow. We denote the unbound fraction of the shocked disk and star material as $f_{\rm ub}^{\rm d}$ and $f_{\rm ub}^\star$, respectively. In what follows, we shall typically leave $f_{\rm ub}^\star$, $f_{\rm ub}^{\rm d} < 1$ as free parameters that will depend primarily on the relative inclination between the disk and the EMRI's orbital plane (see Section \ref{sec:conclusions}).  For example, a retrograde orbit maximizes the collision velocity, $v_{\rm rel}\approx 2v_{\rm k}$, leading to a high ejected fraction, whereas for a nearly aligned orbit, $v_{\rm rel} \ll v_{\rm k}$, and most of the shocked material will remain bound.

Taking $R_{\rm eff}\approx R_\star$, the ratio of shocked star to shocked disk ejecta can be written:
\begin{equation}
    \frac{\delta m_{\star}}{\delta m_{\rm d}} \approx \eta \pfrac{\rtidal}{a_0} \pfrac{\MBH}{m_\star}^{2/3} \approx 
    71 \; \frac{\eta_{0.03} M_{\bullet,6}^{2/3}}{P_{\rm orb,24}^{2/3}  m_1^{0.2}} .
\end{equation}
Ablated stellar material ($\delta m_\star$) thus typically exceeds shocked disk ejecta ($\delta m_{\rm d}$) for most periods $P_{\rm orb} < 600 \, \rm d$ of interest.\footnote{Note however, that Eq.~\eqref{eq:delta_m_abl} is calibrated from simulations spanning a relatively narrow range of collision velocities, corresponding to orbital periods $P_{\rm orb}\lesssim 10 \, \rm d$ \citep{Yao_25}. The stripping efficiency may change considerably for the lower collision velocities at longer orbital periods.} The remaining, bound ejecta, $(1-f_{\rm ub}^{\rm d})\delta m_{\rm d}$ and $(1-f_{\rm ub}^\star)\delta m_\star$, respectively, are likely incorporated back into the disk, though they may (at least temporarily) settle into a hot corona-like layer above and below the midplane (e.g., \citealt{Linial_Metzger_23}; their Appendix C). 

As both $\delta m_\star$ and $\delta m_{\rm d}$ scale with $\Sigmad(a_0)$ (Eq.~\ref{eq:delta_m_disk} and \ref{eq:delta_m_abl}), it shall prove convenient to scale the average outflow rate,
\begin{equation} \label{eq:mdot_gen}
    \dot{m}_{\rm w} = \frac{\delta m_{\rm ub}}{P_{\rm orb}/2} \equiv \xi_{\rm w} \dot{m}_{\rm d}(a_0) \,,
\end{equation}
to the local accretion rate, $\dot{m}_{\rm d}(a_0)=3\pi \Sigmad(a_0) \nu(a_0)$.  Here, $\delta m_{\rm ub} = f_{\rm ub}^\star\delta m_\star + f_{\rm ub}^{\rm d} \delta m_{\rm d}$ is the unbound mass per collision, and we have defined
\begin{equation} \label{eq:xi_w_gen}
    \xi_{\rm w} \equiv \frac{\dot{m}_{\rm w}}{\dot{m}_{\rm d}(a_0)} = \frac{2}{3}\frac{\delta m_{\rm ub}}{\pi a_0^2\Sigmad(a_0)} \frac{t_{\rm v}(a_0)}{P_{\rm orb}} \,,
\end{equation}
where $t_{\rm v}(a_0) \equiv 3\pi a_0^2 \Sigmad / \dot{m}_{\rm d}(a_0)$ is the disk's local viscous time at the collision radius.  In what follows, we normalize $t_{\rm v}$ to $10^{3} \,P_{\rm orb},$ typical of that needed to fit the late-time UV/X-ray TDE light-curves in a spreading disk model (\citealt{Mummery+24}, although some TDEs suggest a much faster viscous time, e.g., \citealt{Guolo_2025_GSN069_time_dependent}).

\begin{figure}
    \centering
    \includegraphics[width=1.05\linewidth]{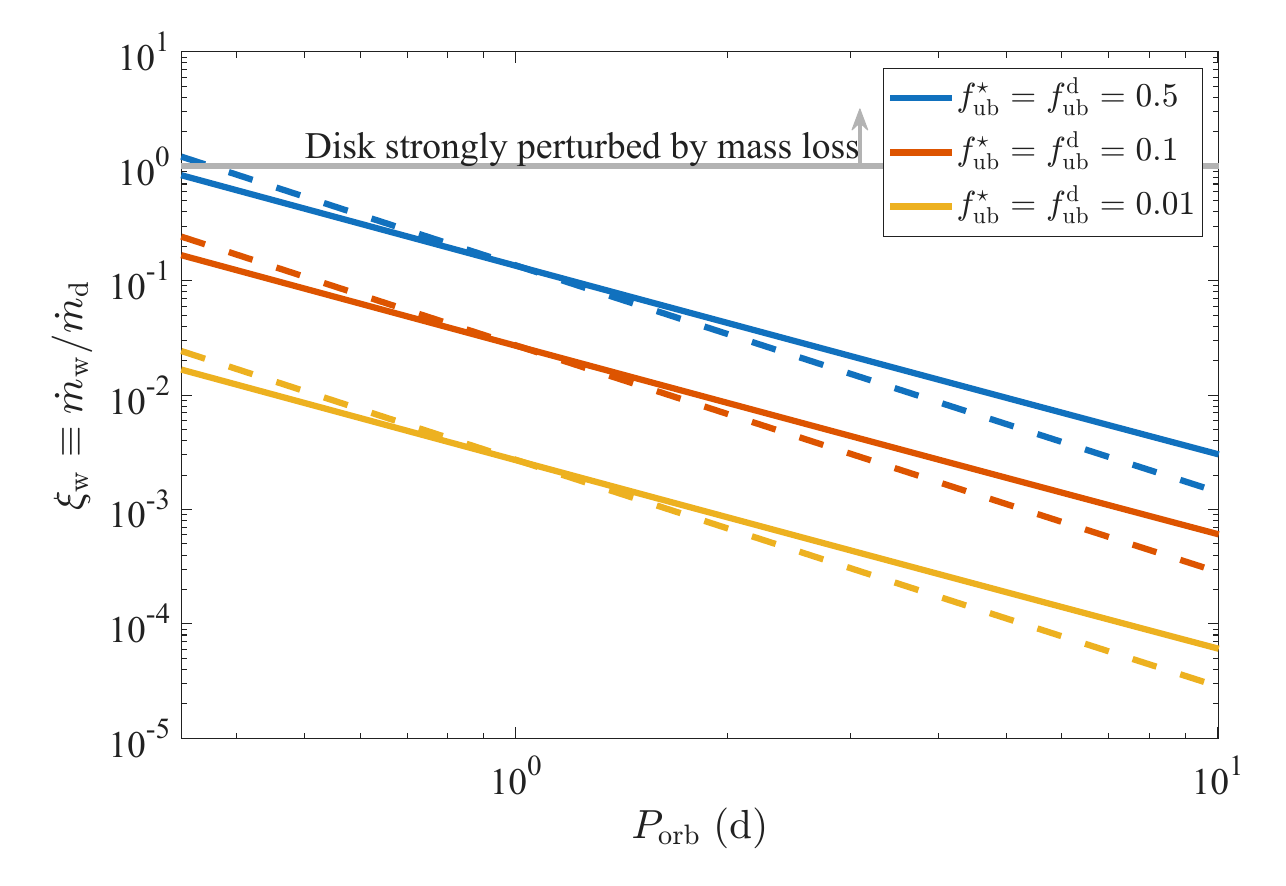}
    \caption{EMRI-disk collision-induced mass outflow rate relative to the local disk accretion rate, $\xi_{\rm w} \equiv \dot{m}_{w}/\dot{m}_{\rm d}$, as a function of EMRI orbital period. Different colors span a plausible range of values of $f_{\rm ub}^\star$ and $f_{\rm ub}^{\rm d}$ (the unbound fraction of the ablated star and shocked disk material, respectively). Solid lines correspond to a constant viscosity, $\nu=10^{19} \, \rm cm^2 \, s^{-1}$, while dashed lines assume $t_{\rm v}/P_{\rm orb} = 10^3$ (\citealt{Mummery&VanVelzen24}). Here $\MBH=10^6 \, \rm M_\odot$, $m_\star=1\,\rm M_\odot$, $\eta=0.03$.}
    \label{fig:xi_w_vals}
\end{figure}

If the mass loss is dominated by shocked disk material, then $\delta m_{\rm ub} \approx f_{\rm ub}^{\rm d} \delta m_{\rm d}$ and hence
\begin{multline} \label{eq:xi_w_disk}
    \xi_{\rm w} \approx \frac{2}{3}\pfrac{R_{\rm eff}}{a_0}^2 \frac{t_{\rm v}}{P_{\rm orb}} \approx \\
    4\times 10^{-3} \, f_{\rm ub}^{\rm d} \pfrac{t_{\rm v}}{10^3P_{\rm orb}} m_1^{1.6} M_{\bullet,6}^{-2/3} P_{\rm orb,24}^{-4/3} \,,
\end{multline}
whereas if ablated stellar material comprises most of the outflow ($\delta m_{\rm ub} \approx f_{\rm ub}^{\star} \delta m_{\star}$), then
\begin{equation} \label{eq:xi_w_abl}
    \xi_{\rm w} \approx \frac{2}{3} \eta\pfrac{\rtidal}{a_0}^3 \frac{t_{\rm v}}{P_{\rm orb}} \approx \\
    0.3 \, f_{\rm ub}^{\rm \star}\pfrac{t_{\rm v}/P_{\rm orb}}{10^3} P_{\rm orb,24}^{-2} m_1^{1.4} \,.
\end{equation}
Fig.~\ref{fig:xi_w_vals} illustrates the dependence of $\xi_{\rm w}$ on $P_{\rm orb}$ for different assumptions regarding the unbound ejecta ($f_{\rm ub}^\star$, $f_{\rm ub}^{\rm d}$) and the disk viscosity prescription (see further discussion in Section \ref{sec:diskevo}), fixing $\eta$, $\MBH$, $m_\star$ at fiducial values.  Across most of the parameter space we expect $\xi_{\rm w} \ll 1 $, although $\xi_{\rm w} \gtrsim 1$ is possible for short orbital periods.  The next section describes the key role played by $\xi_{\rm w}$ in the disk-star evolution.  Given uncertainties in the viscous timescale and the fraction of unbound debris, we consider a wide range of $\xi_{\rm w} \sim 10^{-3}-10$. 

\section{Outflows from the Disk and EMRI}
\label{sec:diskevo}

Having presented the key components of the model, we now examine the interaction and mutual evolution of the EMRI and the spreading disk, which dictate the resulting unbound outflow and its time-dependent mass-loss rate, $\dot{m}_{\rm w}(t)$. This section highlights the different regimes of star-disk interaction and estimates the resulting mass and energy of the outflows.

In focusing on star-disk interactions, we neglect any intrinsic winds or outflows from the disk. Even absent an EMRI, disk outflows may be particularly important at early times after the TDE when the accretion rate is highest (e.g., \citealt{Strubbe&Quataert09,Metzger&Stone16,Price+24,Tuna25}), or at very late times, once the disk transitions to a radiatively inefficient state (e.g., \citealt{Giannios&Metzger11,Tchekhovskoy+14}).

\subsection{Long Term Outcomes of Star-Disk Collisions} \label{sec:outcomes}

\begin{deluxetable*}{llllll}
\tablecaption{Outcome regimes of star--disk encounters and associated outflows\label{tab:regimes_outflows}.}
\tablehead{
\colhead{Regime} &
\colhead{Ejecta} &
\colhead{Key Conditions} &
\colhead{$m_{\rm w}^{(a)}$} &
\colhead{$\Delta t_{\rm w}^{(b)}$} &
\colhead{Outcome Description}
}
\startdata
1) Both survive, disk dominated
& Disk
& $\xi_{\rm w} \ll 1,\; k \approx 1$
& $\xi_{\rm w} m_{\rm d}(a_0)$
& $t_{\rm v}(a_0)$
& Long-lived star--disk collisions (+QPEs?) \\
2) Both survive, EMRI dominated
& EMRI
& $t_{\rm abl} \gg t_{\rm v}(a_0)$
& $m_\star (t_{\rm v}/t_{\rm abl})$
& $t_{\rm v}(a_0)$
& Long-lived star--disk collisions (+QPEs?) \\
3) Depleted disk, EMRI survives
& Disk
& $\xi_{\rm w} \gtrsim 1$
& $m_{\rm d}(a_0)$
& $t_{\rm dep}$
& Disk emission suppressed \\
4) EMRI destroyed, disk survives
& EMRI
& $t_{\rm abl} \ll t_{\rm v}(a_0)$
& $m_\star$
& $t_{\rm abl}$
& QPEs cease alongside outflow ejection \\
\enddata
\tablecomments{$^{(a)}$Total mass of collision ejecta; $^{(b)}$Duration of collision outflows}
\end{deluxetable*}

Mass ejection is determined by the coupled evolution of the star and the impacted disk. As shown in Sec.~\ref{sec:massejection}, the mass outflow rate traces the evolution of $\Sigmad(a_0)$ -- the local disk surface density at the EMRI separation. Mass ejection peaks when the disk’s outer edge coincides with the EMRI, i.e., $a_0\lesssim r_{\rm d} \lesssim 2a_0$. This phase of maximal mass ejection ensues on the local viscous time $t \sim t_{\rm v}(a_0)$ and lasts a comparable duration (for more details, see Sec.~\ref{sec:diskmodel}). Depending on whether the outflow is dominated by ablated star or shocked disk, the total ejecta mass is fundamentally limited, respectively, by the EMRI mass or that remaining in the disk once it has spread to $r_{\rm d} \approx a_0$. 

The ultimate fate of the star-disk system depends on the ratio of the disk's spreading time, $t_{\rm v}$, and either the ablation time of the star, $t_{\rm abl} \approx (m_\star/\delta m_\star) P_{\rm orb}/2$, or the depletion time of the disk, $t_{\rm dep} \approx (m_{\rm d}/\delta m_{\rm d})P_{\rm orb}/2$, respectively. Their ratio can be written:
\begin{multline}
\label{eq:tabl2tdep}
    \frac{t_{\rm dep}}{t_{\rm abl}} = \frac{\delta m_\star/m_\star}{\delta m_{\rm d}/m_{\rm d}} \approx 2^{-1/2}f_{\rm d} \pfrac{\rtidal}{a_0}^{1/2} \frac{\delta m_\star}{\delta m_{\rm d}} \approx \\
    24 \, \eta_{0.03} f_{\rm d} P_{\rm orb,24}^{-1} M_{\bullet,6}^{2/3} m_1^{0.7}\,,
\end{multline}
where for simplicity we assume $m_\star=m_\star^{\rm tde}$. For typical parameters, we see that the star will be substantially ablated before the disk is depleted of its gas ($t_{\rm abl} \ll t_{\rm dep}$), a point we shall return to in Sec.~\ref{sec:Outflow_abl_domninated}.

Table \ref{tab:regimes_outflows} summarizes the 4 qualitative outcomes of star-disk interactions:
\begin{enumerate}
    \item The disk evolves rapidly, leaving both the star and the disk mostly unscathed ($t_{\rm v} \ll t_{\rm dep},t_{\rm abl}$). The outflow is dominated by ejected disk material and stellar ablation is mostly inefficient ($t_{\rm dep} \ll t_{\rm abl}$). 
    \item Same as 1, but the outflow is dominated by ablated stellar material ($t_{\rm abl} \ll t_{\rm dep}$) 
    \item The disk evolves slowly, ablation is inefficient and the star survives, but the disk's mass is significantly diminished by EMRI collisions $(t_{\rm dep} \ll t_{\rm v} \ll t_{\rm abl})$.
    \item The disk evolves slowly and retains most of its mass, but the star is significantly ablated $(t_{\rm abl}\ll t_{\rm v} \ll t_{\rm dep})$.
\end{enumerate}

The next section (Sec.~\ref{sec:diskmodel}) explores these scenarios using a time-dependent disk model, focusing on possibility 1.  We return to scenarios 2, 3 and 4 in Sec.~\ref{sec:Outflow_abl_domninated}.

\subsection{Time dependent disk model}
\label{sec:diskmodel}

The surface density of an axisymmetric height-integrated viscously spreading disk obeys a one-dimensional diffusion equation \citep[e.g.,][]{Pringle81}
\begin{equation} \label{eq:Disk_Equation_Sink}
    \pdiff{\Sigmad}{t} = \frac{3}{r} \pdiff{}{r} \left[ r^{1/2} \pdiff{}{r}\left( \nu r^{1/2} \Sigmad \right) \right] -S(a_0) \,.
\end{equation}
Here, $\nu$ is effective kinematic viscosity -- encapsulating the details of the disk's vertical structure, thermal equilibrium and angular momentum transport. If $\nu \propto r^{p_\nu}$, Eq.~\eqref{eq:Disk_Equation_Sink} permits a well-studied self-similar solution \citep[e.g.,][]{Pringle81,Shen&Matzner14}. Recently, \citet{Alush&Stone25} studied the dynamics of geometrically thin, magnetically elevated accretion disks which cool radiatively, where vertical pressure support is provided by magnetic fields. Under these conditions and assuming alpha-viscosity, i.e., $\nu=\sqrt{G\MBH r} (H_{\rm d}/r)^2 \alpha$ \citep[e.g.,][]{Shakura&Sunyaev73}, they found $\nu \propto r^{5/7} \Sigmad^{2/7}$. Although this functional form of viscosity does not permit a known analytical solution, comparison of the numerical solution to an idealized linear model with $\nu(r)\propto r^{p}$ shows that $p_\nu=0$ (i.e., constant $\nu$, or equivalently $H_{\rm d}\propto r^{3/4}$) captures the time dependence of disk spreading as well as the resulting rate of accretion (albeit a different radial profile of the inner disk surface density; see Fig. 4 of \citealt{Alush&Stone25}). Taking $p_\nu=0$ and calibrating to the TDE circularization radius, $r_{0} \approx 2 \rtidal$, we have
\begin{equation}
    \nu \approx 4\times10^{19} \, {\rm cm^2 \, s^{-1}} \left(\frac{r_{0}}{2\rtidal}\right)^{1/2} M_{\bullet,6}^{1/2} 
    \left(\frac{t_{\rm v}/P_{\rm orb}}{10^{3}}\right)^{-1}.
    \label{eq:viscosity}
\end{equation}
The sink term in Eq.~\eqref{eq:Disk_Equation_Sink}, 
\begin{eqnarray}
    S(a_0)&=& k\frac{\dot{m}_{\rm w}(\Sigmad(a_0))}{2\pi a_0} \delta (r-a_0)  \nonumber \\
    &=& \frac{3k}{2} \xi_{\rm w} \frac{\Sigmad(a_0) \nu(a_0)}{a_0} \delta (r-a_0) \,,
    \label{eq:sink}
\end{eqnarray}
captures the change in disk mass due to collisions with an EMRI localized at radius $r = a_0$, where in the second line we use $\xi_{\rm w}$ from Eq.~\eqref{eq:xi_w_gen}. Although mass is ejected in discrete collisions twice per orbit, we treat the sink as a continuous mass loss term, considering the separation of timescales, $t_{\rm v}\gg P_{\rm orb}$.

The dimensionless parameter $k$ entering Eq.~\eqref{eq:sink} relates the mass outflow rate, $\dot{m}_{\rm w}$, to the net change in the local disk mass due to ejection from EMRI collisions (mass loss)  and the incorporation of ablated EMRI material (mass gain). Neglecting ablation ($\eta=0$), outflows 
arise purely from the unbound portion of the excavated disk mass, and $k=1$. However, values $k<1$ occur when stellar ablation is significant and even $k<0$ is permitted, in which case $S(a_0)$ can become a source, rather than a sink term\footnote{For example, $k<0$ occurs for $(1-f_{\rm ub}^\star)\delta m_\star \gtrsim f_{\rm ub}^{\rm d} \delta m_{\rm d}$, and if the bound ablated material (i.e., $(1-f_{\rm ub}^\star)\delta m_\star$), is ultimately incorporated into the disk. This counterintuitive limit, in which the disk mass \textit{grows} despite outflows being launched (e.g., \citealt{Linial_Metzger_24b}), may occur when the same collisions that expel mass, also efficiently ablate the EMRI and locally increase the disk mass with stellar debris.}. For concreteness in what follows, we shall assume $k=1$.

\subsubsection{Disk Evolution}
We solve Eq.~\eqref{eq:Disk_Equation_Sink} numerically using a time-implicit Crank-Nicolson scheme. For the initial state of the bound TDE debris, we assume a localized ring of mass $m_{\rm 0}$ and radius $r_{\rm 0}$, with a Gaussian radial surface density profile of thickness $0.05 \, r_0$.  Although we assume that all of the bound mass has the same specific angular momentum, such that $r_0=2\rp \lesssim 2 \rtidal$, the disk evolution at late times of interest is largely insensitive to details of the initial mass distribution. For the inner boundary condition we take $\partial\ln\Sigmad/\partial \ln r=-\partial\ln{\nu}/\partial\ln{r}$, at $r_{\rm in}=10^{-2} \, r_0$, which enforces radially constant accretion through the inner regions of the disk (i.e., $\dot{m}_{\rm d}(r) \simeq 3\pi \nu \Sigmad = \rm const$).  We assume a constant viscosity $\nu=\nu_0$ (i.e., $p_\nu=0$), with the viscous time $t_{\rm v,0} = r_{\rm 0}^2/\nu_0$ at the initial ring radius serving as the simulation unit of time.  

\begin{figure}
    \centering
    \includegraphics[width=1.0\linewidth]{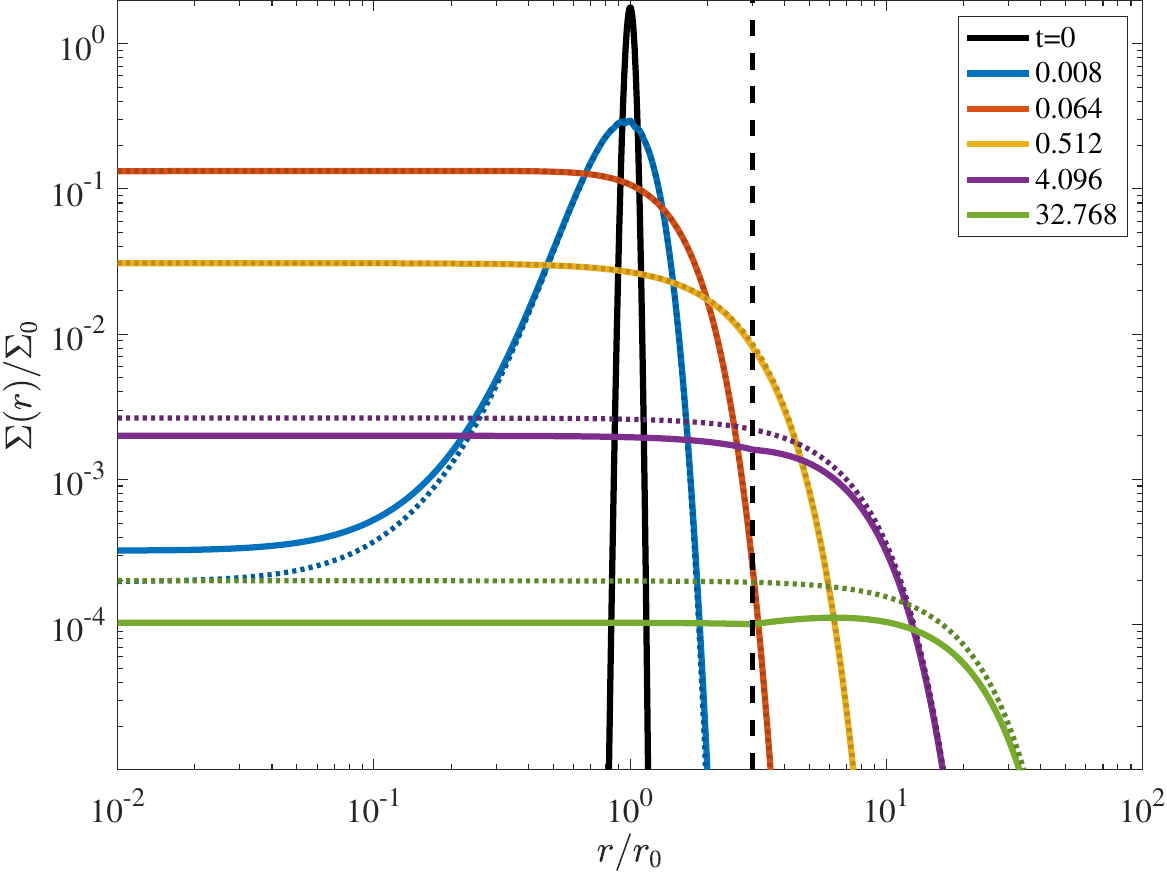}
    \caption{Spreading disk surface density evolution, including the effects of mass-loss from star-disk collisions. A narrow Gaussian ring of mass centered at $r_0$ begins to viscously spread at time $t=0$ (solid black line). Solid lines show the disk evolution in the presence of a weak sink term at $a_0=3r_0$ due to star-disk collisions (vertical black dashed line). The disk evolution in the absence of a sink term is shown as dotted lines (the analytical Green's function, Eq.~\eqref{eq:Sigma_r_t_Green}). Time is measured in units of $t_0=r_0^2/\nu_0$, where $\nu_0$ is the (constant) kinematic viscosity.}
    \label{fig:SpreadingDisk_a0=3_xiw=0.6}
\end{figure}

Fig.~\ref{fig:SpreadingDisk_a0=3_xiw=0.6} shows snapshots in the evolution of the disk surface density for an EMRI located at $a_0=3r_0 = 6 \rtidal$ on an orbit of period $P_{\rm orb}(a_0) \approx 1.7 \, {\rm d} \,(a_0/3r_0)^{3/2} (m^{\rm tde}_{\star,1})^{0.7}$.  Solid lines show a model for which EMRI collisions are neglected ($\xi_{\rm w}=0$).  The Green's function solution to the diffusion equation for a constant viscosity disk with no sink term is given by \citep{LyndenBell_74},
\begin{multline} \label{eq:Sigma_r_t_Green}
    \Sigmad^{\rm 0}(r,t) = \frac{m_0}{12 \pi \nu_0t} \pfrac{r}{r_0}^{-1/4} \\
    \exp\left( -\frac{r^2+r_0^2}{12\nu_0t}\right) I_{1/4}\pfrac{rr_0}{6\nu_0t} \,,
\end{multline}
where $I_{1/4}$ is the modified Bessel function. This agrees well with our model neglecting disk mass loss ($\xi_{\rm w}=0$; dotted line in Fig.~\ref{fig:SpreadingDisk_a0=3_xiw=0.6}) except at very early times ($t=0.008 \, t_0$), due to the initial ring being of finite radial thickness instead of a delta function.

Solid lines in Fig.~\ref{fig:SpreadingDisk_a0=3_xiw=0.6} show the disk evolution, including the effects of mass-loss from stellar collisions, $\xi_{\rm w}=0.6$ (Eq.~\eqref{eq:xi_w_gen}). The earliest phases of the disk's evolution closely match that neglecting mass-loss ($\xi_{\rm w} = 0$). However, once the disk expands to reach the EMRI orbit at $a_0=3r_0$ (vertical dashed line), outflows ensue and the remaining disk mass is greatly reduced compared to the unperturbed case (Eq.~\eqref{eq:m_disk_at_a0}).

\subsubsection{Outflow rate and total mass} \label{sec:Outflow_k=1}

\begin{figure}
    \centering
    \includegraphics[width=1.0\linewidth]{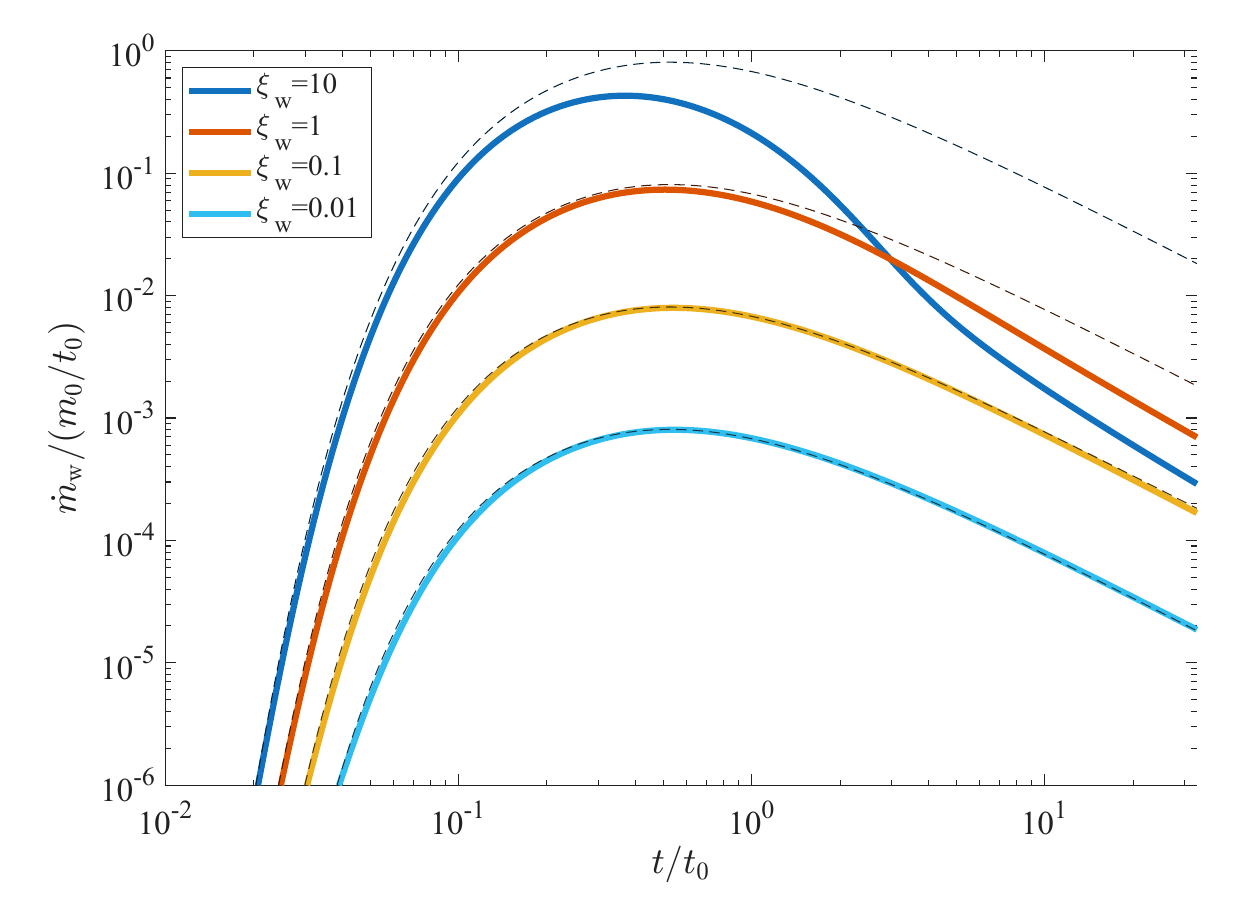}
    \includegraphics[width=1.05\linewidth]{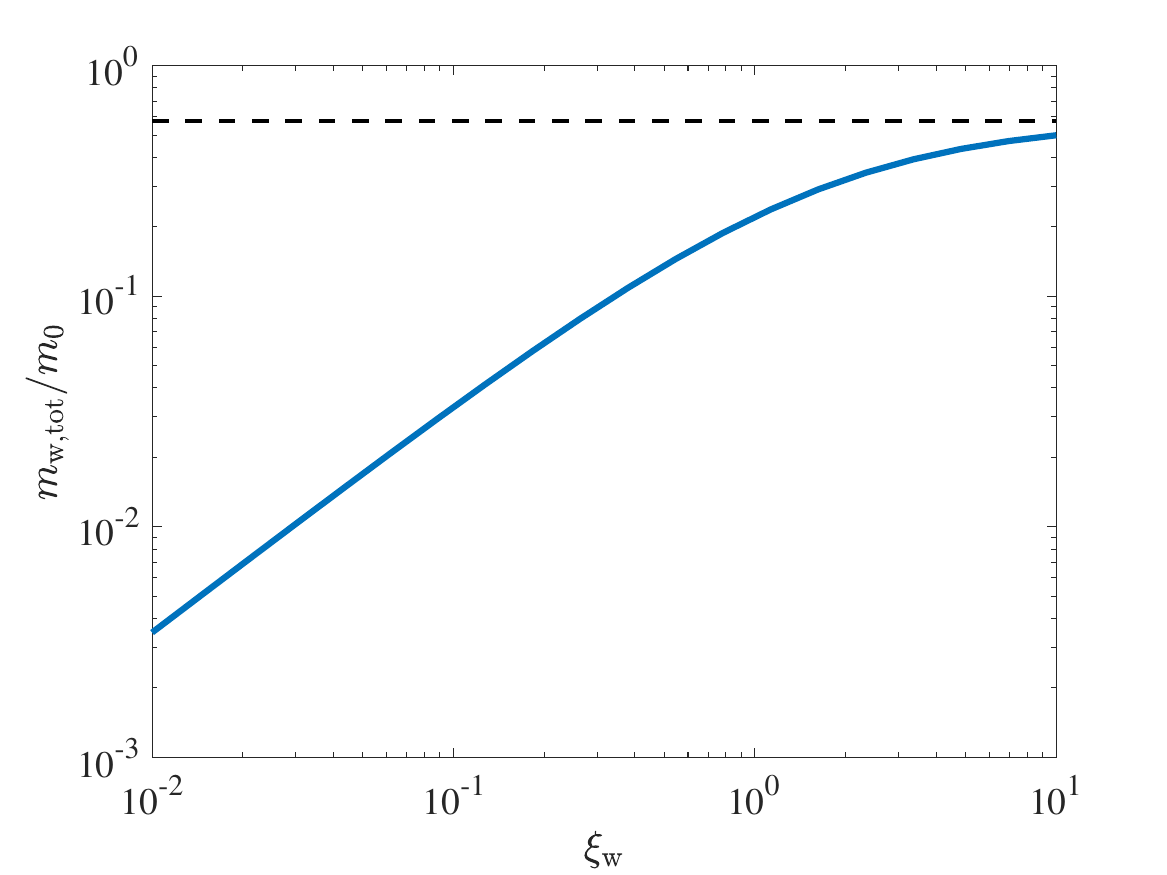}
    \caption{Top: Time evolution of the outflow ejection rate $\dot{m}_{\rm w}$ for different values of the dimensionless mass-loss efficiency $\xi_{\rm w}$ (Eq.~\eqref{eq:xi_w_gen}) indicated in the legend. All models assume $k=1$, corresponding to an outflow dominated by ejected disk material and negligible stellar ablation. Dashed lines show the estimated mass loss one obtains, neglecting the back reaction of collisional mass-loss on the disk evolution (Eq.~\eqref{eq:mdot_out_unperturbed_disk}). Bottom: The total integrated outflow mass as a function of $\xi_{\rm w}$. The total ejection is limited to $m_0(a_0/r_0)^{-1/2}$ (dashed horizontal line) -- the remaining disk mass once its outer radius approaches the EMRI orbit at $a_0$.}
    \label{fig:Mdot_outflow_xi_w}
\end{figure}

Fig.~\ref{fig:Mdot_outflow_xi_w} shows the time evolution of the wind outflow rate, $\dot{m}_{\rm w}(t)$, for different values of $\xi_{\rm w}\in (0.01,10)$. For $\xi_{\rm w}\ll 1$, the disk is mostly unperturbed, evolving similarly to the $\xi_{\rm w}=0$ case. In this regime, the outflow traces directly the surface density evolution at $r=a_0$ for the unperturbed Green's function (Eq.~\ref{eq:Sigma_r_t_Green}), with
\begin{equation} \label{eq:mdot_out_unperturbed_disk}
    \dot{m}^0_{\rm w}(t) \approx 3\pi \xi_{\rm w} \nu_0 \Sigmad^0(a_0,t) \,.
\end{equation}
This expression, shown with dashed lines in Fig.~\ref{fig:Mdot_outflow_xi_w}, agrees well with the numerical results for $\xi_{\rm w}\ll 1$.  In the limit $a_0\gg r_0$, the outflow rate peaks when $\Sigmad^0(r=a_0,t)$ reaches its maximal value at time $t_{\dot{m}_{\rm max}} = a_0^2/(15\nu_0)$.  Although this is notably faster than the naive viscous time, $t_{\rm v}=a_0^2/\nu_0$, the peak in $\dot{m}_{\rm w}$ is gradual, such that the bulk of the mass ejection still occurs over $\sim t_{\rm v}$: $(\int^{t_{\rm v}} \dot{m}_{\rm w} dt)/(\int^\infty \dot{m}_{\rm w} dt) \approx 0.42$, whereas $(\int^{t_{\dot{m}_{\rm max}}}\dot{m}_{\rm w} dt)/(\int^\infty \dot{m}_{\rm w} dt) \approx 0.05$.

On the other hand, for $\xi_{\rm w} \gtrsim 1$, the disk is strongly perturbed by mass and angular momentum loss through the sink term, leading to its rapid depletion over $t_{\rm v}(a_0)$. At late times, the lower remaining disk mass decreases the outflow rate to well below the unperturbed rate, $\dot{m}^0_{\rm w}(t)$. 

The bottom panel of Fig.~\ref{fig:Mdot_outflow_xi_w} shows the total ejecta $m_{\rm w,tot} = \int \dot{m}_{\rm w}(t) dt$ as function of $\xi_{\rm w}$. As expected, for $\xi_{\rm w}\ll 1$, ${m}_{\rm w,tot}\propto \xi_{\rm w}$, saturating around $\xi_{\rm w}\approx 1$ to the remaining disk mass when its outer edge intercepts the EMRI at $a_0$ (Eq.~\eqref{eq:m_disk_at_a0}; horizontal dashed line). 

For $k \approx 1$ and $\xi_{\rm w} \ll 1$, the ejecta mass, $m_{\rm w,tot}^{\rm d}$, can be estimated by noting that as the disk spreads past the ejection site at $r=a_0$ a fraction $\xi_{\rm w}\ll1$ of the remaining disk (Eq.~\eqref{eq:m_disk_at_a0}) is ejected as an outflow:
\begin{multline} \label{eq:m_w_tot_numerical}
    m_{\rm w,tot}^{\rm d} \approx 0.6 \, \xi_{\rm w} (f_{\rm d} m_{\rm \star}^{\rm tde}/2) (a_0/2\rtidal)^{-1/2} \approx \\ 
    0.2 \, {\rm M_\odot} \, \xi_{\rm w} f_{\rm d}m_{\star,1}^{\rm tde}  P_{\rm orb,24}^{-1/3} \,.
\end{multline}
Assuming the outflow is dominated by ejected disk material ($f_{\rm ub}^{\rm d} \delta m_{\rm d} \gg f_{\rm ub}^{\star} \delta m_{\star}$), and substituting Eq.~\eqref{eq:xi_w_disk} for $\xi_{\rm w}$ into Eq.~\eqref{eq:m_w_tot_numerical}, we obtain:
\begin{multline}
\label{eq:m_w_tot_disk}
    m_{\rm w,tot}^{\rm d} \approx 
    10^{-3} \, {\rm M_\odot} \frac{f_{\rm d} f_{\rm ub}^{\rm d}(t_{\rm v}/10^3P_{\rm orb}) m_1^{1.6} m_{\star,1}^{\rm tde}}{P_{\rm orb,24}^{5/3} M_{\bullet,6}^{2/3}} \,.
\end{multline}

For $k<1$, which allows for the entrainment of ablated stellar material back into the disk, the same qualitative result of Eq.~\eqref{eq:m_w_tot_numerical} applies (albeit with a slightly different prefactor), even if the ejecta is dominated by ablated stellar material, rather than disk material -- the reason is that both $\delta m_{\rm d}$ and $\delta m_\star$ are proportional to $\Sigmad$ (Eq.~\eqref{eq:delta_m_disk} and \eqref{eq:delta_m_abl}). The above estimates break down, however, if the EMRI is destroyed by the disk on a shorter timescale than the disk can spread past it. We discuss this regime qualitatively in the next Section.

\subsection{Outflows dominated by ablated star and EMRI destruction} \label{sec:Outflow_abl_domninated}

Across much of the parameter space, $\xi_{\rm w} \ll 1$ (Fig.~\ref{fig:xi_w_vals}) and hence the spreading TDE disk is not substantially depleted by stellar collisions.  However, even when $\xi_{\rm w} \gtrsim 1$, the impacting EMRI may be destroyed (or at least substantially depleted) by ablation before excavating a significant fraction of the disk's mass via collisions because $t_{\rm abl} < t_{\rm dep}$ (Eq.~\eqref{eq:tabl2tdep}).  In fact, disk depletion ($\xi_{\rm w} \gtrsim 1$) by an EMRI that self-consistently survives the same interaction ($t_{\rm abl} \gg t_{\rm dep}$) is possible only for either very long orbital periods, $P_{\rm orb}\gg 20 \, {\rm d} \, f_{\rm d} M_{\bullet,6}^{2/3}$, or for much lower ablation efficiencies, $\eta \ll 0.03$, than found by \citet{Yao_25}.

Whether the EMRI can be substantially ablated by disk collisions depends on the ratio of $t_{\rm abl}$ to the viscous time over which the disk spreads past it, 
\begin{multline} \label{eq:t_abl_over_t_v}
    \frac{t_{\rm abl}}{t_{\rm v}(a_0)} \approx \frac{m_\star/2\delta m_{\star}}{t_{\rm v}/P_{\rm orb}} \approx \\
    3.6 \; (f_{\rm d} \eta_{0.03})^{-1} P_{\rm  orb,24}^{7/3} (t_{\rm v}/10^3 \, P_{\rm orb})^{-1} m_1^{-0.4} (m_{\star,1}^{\rm tde})^{-1.2}.
\end{multline}
Here, we have used $\Sigmad^{\rm max} \approx m_{\rm d}(r_{\rm d}=a_0)/a_0^2$ for the peak surface density encountered by the EMRI as the disk's outer edge sweeps past it; this approximation is valid for $a_0 > 2\rtidal^{\rm tde}$, such that the TDE disk forms within the EMRI orbit.

Ablation is thus rapid ($t_{\rm abl} \ll t_{\rm v}$) for sufficiently short orbital periods, 
\begin{equation}
    P_{\rm orb} \lesssim 14 \, {\rm hr} \; (t_{\rm v}/10^3 \,P_{\rm orb})^{3/7} (f_{\rm d} \eta_{0.03})^{3/7} m_1^{0.17} (m_{\star,1}^{\rm tde})^{0.53}, \,
\end{equation}
for which the star loses most of its mass before the disk's outer radius has doubled.\footnote{This estimate would change if a fraction of $\delta m_\star$ is added back to the disk, modifying its structure ($k<1$); however, any increase in the local $\Sigmad$ would only act to accelerate the EMRI's destruction (Eq.~\eqref{eq:delta_m_abl}).}  This limit where $t_{\rm abl}\ll t_{\rm v}$ (and where $\delta m_\star/m_\star\gtrsim \delta m_{\rm d}/m_{\rm d}$) corresponds to outcome 4 outlined in Section \ref{sec:outcomes}.  In this case the outflow is effectively launched in a singular shell of mass $m_{\rm w,tot}\approx f_{\rm ub}^\star m_\star$, released over the time interval $(t_{\rm v},t_{\rm v}+t_{\rm abl})$, with a typical velocity $v_{\rm w} \approx v_{\rm k}$. See further discussion on survivability of QPE systems in Sec. \ref{sec:TDE_QPE_Survivability}.

In the complementary limit, $t_{\rm abl} \gg t_{\rm v}$ (e.g., longer $P_{\rm orb}$ or 
smaller $t_{\rm v}/P_{\rm orb}$),
the star loses only a small fraction of its mass, $\sim m_\star (t_{\rm v}/t_{\rm abl})$ during the first viscous time after $r_{\rm d} \approx a_0$. At times $t\gtrsim 2t_{\rm v}$, $\dot{m}_{\rm w} \propto \Sigmad(t;a_0)\propto t^{-5/4}$, such that most of the total ejecta mass occurs over the first viscous time after the disk reaches the orbiter. Thus, for ablation-dominated ejecta, we have
\begin{multline} \label{eq:m_w_tot_star}
    m_{\rm w,tot}^{\rm \star} \approx \dot{m}_{\rm w}(t_{\rm v})t_{\rm v}\approx f_{\rm ub}^\star m_\star \pfrac{t_{\rm v}}{t_{\rm abl}} \approx \\
    0.3 \, {\rm M_\odot} f_{\rm ub}^\star f_{\rm d} \eta_{0.03} P_{\rm orb,24}^{-7/3} (t_{\rm v}/10^3 \,P_{\rm orb}) m_1^{1.6} 
    \,,
\end{multline}
but limited to a maximum value $m_{\rm w,tot} \lesssim f_{\rm ub}^\star m_\star$. In the final line, we have assumed $m_\star\approx m_\star^{\rm tde}$.

\subsection{Summary of Disk-Star Collision Outflows}

To summarize, disk-star collisions result in an outflow of characteristic velocity $v_{\rm w}\approx v_{\rm k}$. Depending on the unbound fraction of the intercepted disk mass or the ablated stellar debris (i.e., $f_{\rm ub}^{\rm d}$ and $f_{\rm ub}^\star$), the outflow may be dominated by either disk or EMRI material.  For long EMRI periods the disk is largely unperturbed and the collision-induced outflow is dominated by disk material ($m_{\rm w,tot}^{\rm d}$; Eq.~\ref{eq:m_w_tot_disk}; Sec.~\ref{sec:Outflow_k=1}), while for shorter periods the ejecta can be dominated by partial or full destruction of the stellar EMRI via ablation ($m_{\rm w,tot}^{\rm \star}$; Eq.~\ref{eq:m_w_tot_star}). Depending on the dominant source of the outflow, approximately $m_{\rm w,tot} \approx (10^{-3}-1) \, \rm M_\odot$ are ejected in total.

\begin{figure}
    \centering
    \includegraphics[width=\linewidth]{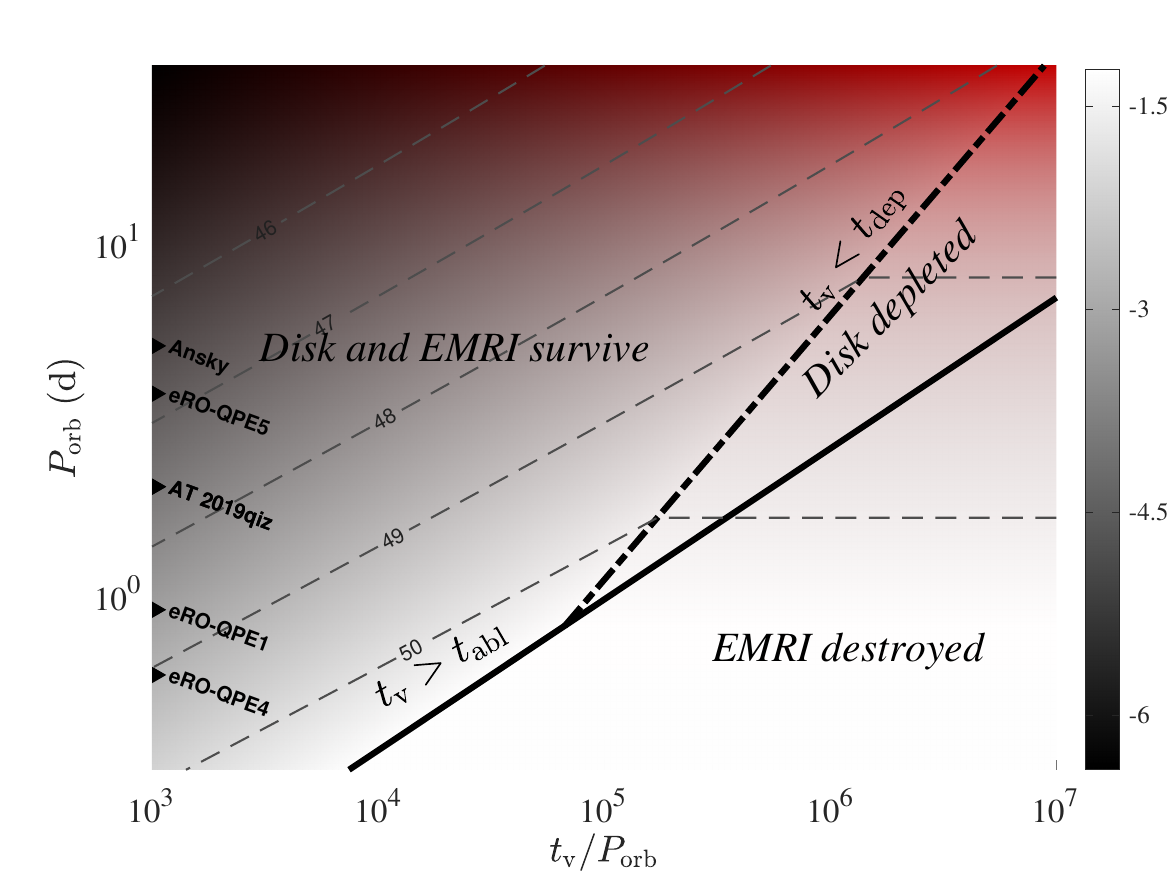}
    \caption{Regimes of star-disk interaction outcomes, as a function of orbital period and the disk's viscous time. Here we assume $f_{\rm d}=0.1$, $f_{\rm ub}^\star=0.1$, $f_{\rm ub}^{\rm d}=0.5$, and $m_\star=m_\star^{\rm tde}=0.5 \, \rm M_\odot$. Below the solid black line, the ablation time is shorter than the disk spreading time, and the EMRI is destroyed (and the disk survives). Below the dash-dotted line, the disk depletion time is shorter than $t_{\rm v}$ and the disk is mostly depleted (and the EMRI survives). The color map intensity (dark-bright) represent the outflow mass, as indicated in the color bar, $\log_{\rm 10}(m_{\rm w,tot}/\rm M_\odot)$. The colormap saturation indicates the contribution of disk vs. EMRI mass to the outflow (redder color means relatively more disk mass). Dashed contours show the outflow kinetic energy, $\log_{10}(E_{\rm w}/\rm erg)$. Orbital periods of a few known QPE sources are shown on the left.}
    \label{fig:OutcomesRegimes}
\end{figure}

Fig.~\ref{fig:OutcomesRegimes} demonstrates different outcomes of disk and EMRI interaction, as a function of $P_{\rm orb}$ and $t_{\rm v}/P_{\rm orb}$, assuming $f_{\rm d}=0.1$, $f_{\rm ub}^\star=0.1$, $f_{\rm ub}^{\rm d}=0.5$, and $m_\star=m_\star^{\rm tde}=0.5 \, \rm M_\odot$. The EMRI ablation time is shorter than the disk's spreading time in the bottom right corner, below the black solid line, representing the regime where the EMRI is destroyed (or at least substantially stripped) within $t_{\rm abl}$. The total ejecta mass is dominated by the EMRI, and equals roughly $f_{\rm ub}^\star m_\star$ (outcome 4 in Sec.~\ref{sec:outcomes}). In the wedge below the dash dotted line and above the solid line, $t_{\rm dep} < t_{\rm v}$ and the disk is significantly depleted before the disk spreads, with a total mass set by the remaining disk mass, $f_{\rm ub}^{\rm d} m_{\rm d}(a_0)$ (outcome 3 in Sec.~\ref{sec:outcomes}). Across the rest of the parameter space, the disk spreads fast enough past the EMRI that both survive the most destructive phase of evolution, when $r_{\rm d} \approx a_0$. The color tone indicates the total ejecta mass, $m_{\rm w,tot}$ (see color bar, indicating $\log_{10}(m_{\rm w,tot}/\rm M_\odot)$), while redder regions indicate higher $m_{\rm w,d}/m_{\rm w,\star}$.  From the timescale ratio in Eq.~\eqref{eq:tabl2tdep}, we see that the ejecta is dominated by ablated stellar material up to $P_{\rm orb} \approx 50 \, \rm d$ (outside the shown range) for the fiducial parameters assumed here.

\subsection{Connection to observable quantities }

In the previous sections we estimated the outflow properties ($M_{\rm w}$, $v_{\rm w}$, $\Delta t_{\rm w}$) in terms of the properties of the accretion disk and the EMRI orbit (e.g., $P_{\rm orb}$ and $m_\star$, $m_{\rm d}$, $\nu$, etc.). However, as the latter are not generally directly measurable, here we describe how to invert these relations to obtain the underlying EMRI/disk properties in terms of the system observables.

Given an observed outflow velocity $v_{\rm w}/c = 0.05v_{0.05}$, the EMRI orbital period can be estimated from Eq.~\eqref{eq:v_k_from_P}:
\begin{equation}
    P_{\rm orb} = 2\pi \frac{G\MBH}{v_{\rm w}^3} \approx 2.9 \, {\rm d} \,M_{\bullet,6} v_{0.05}^{-3} \,,
\end{equation}
where here and in what follows we assume $v_{\rm w} \sim v_{\rm k}$ appropriate for a typical star-disk collision angle. 

The outflow launching time is directly related to the viscous time at the EMRI separation, $\Delta t_{\rm w} \approx t_{\rm v}(a_0) = a_0^{2}/\nu$.  Thus we can write 
\begin{eqnarray}
 \frac{P_{\rm orb}}{t_{\rm v}} = \frac{G\MBH}{v_{\rm w}^3 \Delta t_{\rm w}} \approx  1.3\times10^{-3} \, M_{\bullet,6} v_{0.05}^{-3} \Delta t_{\rm w,yr}^{-1},
 \label{eq:viscosity_obs2}
\end{eqnarray}
where $\Delta t_{\rm w,yr}=\Delta t_{\rm w}/\rm yr$.

Lastly, the outflow mass is approximately
\begin{equation}
    m_{\rm w} \approx \frac{2E_{\rm k}}{v_{\rm w}^2} \approx 4.5\times 10^{-2} \, {\rm M_\odot} \, E_{\rm 50} v_{0.05}^{-2} \,.
    \label{eq:mw_obs}
\end{equation}
If dominated by shocked disk material, the ejecta mass $m_{\rm w}$ should not exceed $m_{\rm d}(a_0)$ (Eq.~\eqref{eq:m_disk_at_a0}), i.e.
\begin{equation}
    m_{\rm w} <m_{\rm d}(a_0) \lesssim 0.24 \,{\rm M_\odot} \, f_{\rm d} (m_{\star,1}^{\rm tde})^{1.2} v_{0.05} M_{\bullet,6}^{-1/3} \,,
\end{equation}
thus setting a condition
\begin{equation}
    E_{50} v_{0.05}^{-3} M_{\bullet,6}^{1/3} f_{\rm d} ^{-1} (m_{\star,1}^{\rm tde})^{-1.2} < 5.4 \,.
\end{equation}

If the outflow is instead dominated by ablated stellar material, the ejecta mass is limited to the EMRI mass, $m_{\rm w} < m_\star$. Combining Eqs.~\eqref{eq:m_w_tot_star}, \eqref{eq:viscosity_obs2}, we find
\begin{equation}
    m_{\rm w,tot} \approx 10^{-2} \, {\rm M_\odot} \,f_{\rm ub}^\star f_{\rm d} \eta_{0.03} v_{0.05}^2 \Delta t_{\rm w,yr} m_1 \,,
\end{equation}
and so from Eq.~\eqref{eq:mw_obs} we can relate the observed quantities to the ablation parameters: 
\begin{equation}
    E_{\rm 50} v_{0.05}^{-4} \Delta t_{\rm w,yr}^{-1} \approx 0.2 \, f_{\rm ub}^\star f_{\rm d} \eta_{0.03} m_1 \,.
\end{equation}
Here, a slower outflow corresponds to a longer launching delay, $\Delta t_{\rm w}$, since $v_{\rm w}\propto P_{\rm orb}^{-1/3}$ and $\Delta t_{\rm w}$ increase with $P_{\rm orb}$.

\section{Interaction with the Ambient Medium and Delayed Radio Flare}
\label{sec:radio}

The previous Sections have demonstrated that repeated star-disk collisions in the aftermath of a TDE can eject a sizable fraction of the EMRI/disk mass as an outflow of typical velocity $v_{\rm w} \approx 0.02-0.1 \, c$, launched with a delay $\Delta t_{\rm w} \approx t_{\rm v}(a_0) \sim \rm yrs$ after the TDE, carrying a total mass $\sim 10^{-3}-1\,\rm M_\odot$.  We now discuss the radio emission produced as this outflow interacts with gas on larger scales surrounding the SMBH.

The synchrotron ``afterglow'' emission from ejecta propagating into an external medium has been studied extensively over the past decades in the context of a wide variety of transients, including supernovae, neutron star mergers, gamma-ray bursts, and TDEs, spanning diverse regimes of ejecta energy, velocity, and collisionless shock physics (see \citealt{Metzger+15b} for a summary). Here we do not aim to advance this body of theory, nor to provide detailed predictions of the resulting radio light curves or spectra. Instead, we apply canonical results to estimate the approximate properties of the resulting radio flares.


Ejecta of velocity $v_{\rm w} = 0.1v_{w,-1}c$ and kinetic energy $E_{\rm k} = E_{51}10^{51}~{\rm erg}$ propagating into an external medium of constant density $n_{\rm ext} = n_310^3~{\rm cm^{-3}}$, decelerates on the timescale,
\begin{equation} \label{eq:t_dec}
    t_{\rm dec} \simeq 
    \left(\frac{3 m_{\rm w}}{4 \pi n_{\rm ext} m_p v_w^3}\right)^{1/3}
    \approx 3.3~{\rm yr}\,
    E_{51}^{1/3} n_3^{-1/3} 
    \beta_{\rm w,-1}^{-5/3} \,,
\end{equation}
and radial scale $r_{\rm st} \approx v_{\rm w} t_{\rm dec} \approx 3\times 10^{17} {\, \rm cm} \, E_{51}^{1/3} n_{3}^{-1/3} \beta_{\rm w,-1}^{-2/3}$. Although the gaseous density profiles surrounding TDEs are found to decline steeply with radial distance (e.g., $n_{\rm ext} \propto r^{-\gamma}$, with $\gamma \approx1-2.5$; \citealt{Alexander+20}), the uniform ambient density ($\gamma=0$) assumed in Eq.~\eqref{eq:t_dec} remains a good approximation provided one chooses $n_{\rm ext} = n(r_{\rm st})$ (to within a factor $(3/(3-\gamma))^{1/3} \lesssim 1.8$ for $\gamma\lesssim2.5$). At times $t\gg t_{\rm dec}$, the blast wave enters the Sedov-Taylor phase, with $v_{\rm sh}(t) \approx v_{\rm w} (t/t_{\rm dec})^{(\gamma-3)/(5-\gamma)}$ and $r_{\rm sh} \approx r_{\rm st} (t/t_{\rm dec})^{2/(5-\gamma)}$.

As the outflow expands, the synchrotron self-absorption (SSA) frequency decreases with time. The ejecta becomes optically-thin to SSA at an observed frequency $\nu_{\rm obs}$ on a timescale \citep[e.g.,][]{Matsumoto&Piran21}
\begin{equation} \label{eq:t_SSA}
    t_{\rm SSA}(\nu_{\rm obs}) = 11 \, {\rm yr} \, E_{51}^{0.49} n_3^{0.5} \epsilon_{\rm B,-2}^{0.52} \epsilon_{e,-1}^{0.47} \nu_{\rm obs,1\,GHz}^{-1.5} \,,
\end{equation}
where this expression is valid in the limit $t_{\rm SSA} > t_{\rm dec}$ and we have made the standard assumptions that a fraction $\epsilon_{\rm B}$, $\epsilon_e$ of the shock energy is placed into magnetic fields and non-thermal electrons, assuming the latter follow power-law distribution of energies $dN/d\gamma_e \propto \gamma_e^{-p}$. In the above and what follows, we take $p=2.5$. Furthermore, we assume that $\beta_{\rm w} \lesssim 0.3$, such that the deep-Newtonian regime is always applicable \citep[e.g.,][]{Matsumoto&Piran21}.

As long as the the density profile is shallow enough, $\gamma < 12/(p+5) \approx 1.6$, the observed radio light curve will peak at time
\begin{equation}
    t_{\rm pk} \simeq \max(t_{\rm dec},\, t_{\rm SSA}) \,, \; \; \gamma<12/(p+5) \,.
    \label{eq:tpeak}
\end{equation}
Namely, the peak is set by the later of the two epochs:
(1) the transitions from self-absorbed to optically thin emission at the observing frequency, $\nu_{\rm obs}$; or (2) the outflow deceleration time, after a mass comparable to its own is swpet (e.g., \citealt{Nakar_Piran_2011,BarniolDuran&Piran13,Matsumoto&Piran21}). For steeper profiles, $\gamma > 12/(p+5)$, the optically thin emission is already declining before significant deceleration, and the lightcurve instead peaks at $t_{\rm pk} \approx t_{\rm SSA}$.

In dense nuclear environments, where the optical depth remains high for years following ejection, the light-curve rise at GHz frequencies is expected
to be limited by SSA transparency (i.e., $t_{\rm SSA} > t_{\rm dec}$), as observed in several delayed TDE radio
flares such as AT~2019azh \citep{Sfaradi+22} and WTP14adeqka \citep{Golay+25}.  In this regime the light curve peaks at the SSA frequency on the timescale $t \sim t_{\rm SSA}$, at a flux density
\citep[e.g.,][]{Matsumoto&Piran21}
\begin{multline}
\label{eq:Fnu1}
    F_{\nu,\rm obs}^{\rm max} =
    220 \, {\rm mJy} \, \epsilon_{e,-1}^{0.26}  \epsilon_{\rm B,-2}^{0.33} E_{51}^{0.84} \nu_{\rm obs,1\,\rm GHz}^{0.84} d_{100}^{-2} \,,
\end{multline}
where $d=100 \, {\rm Mpc} \, d_{\rm 100}$ is the source distance.  At a fixed observer frequency, the flux rises as $F_\nu \propto t^{1.1}$ at $t<t_{\rm SSA}$ and falls as $F_\nu \propto t^{-1.05}$ at $t>t_{\rm SSA}$ for a constant density medium. If the external profile is steeper, e.g. $n_{\rm ext}\propto r^{-2}$, then both the rise and fall are faster, as $F_\nu \propto t^{1.5}$, and $F_{\nu} \propto t^{-1.5}$, respectively. We stress that all temporal scalings discussed here are with respect to $t$, measured from the \textit{launching time of the outflow}, i.e., offset by $\Delta t_{\rm w}$ from the tidal disruption.

In lower-density environments, or for larger ejecta masses, we instead have $t_{\rm SSA} < t_{\rm dec}$ and the radio peak instead traces the deceleration time, provided that $\gamma < 12/(p+5)$. In this case the peak flux is achieved at $t_{\rm pk} \approx t_{\rm dec}$
\begin{multline}
\label{eq:Fnu2}
    F_{\rm \nu,obs}^{\rm max}(\nu_{\rm obs}) \approx \\
    1.8 \, {\rm mJy} \, \epsilon_{e,-1}  \epsilon_{\rm B,-2}^{0.88} n_0^{0.88} E_{51}^{1.0}
    \beta_{\rm w,-1}^{1.75} \nu_{\rm obs,1 \, GHz}^{-0.75} d_{100}^{-2} \,,
\end{multline}
where here we have rescaled the external density to a lower value $n_{\rm ext}= 1 \,n_0\,\rm cm^{-3}$.  In this regime, and for $\gamma=0$, the rise to peak scales as $F_{\nu} \propto t^{3}$ ($t_{\rm SSA} < t < t_{\rm dec}$), followed by a decay $t^{-1.6}$ at times $t\gg t_{\rm dec}$.

The timescale ratio scales as $t_{\rm dec}/t_{\rm SSA} \approx 0.33 \, E_{51}^{-0.16} n_3^{-0.83} \beta_{\rm w,-1}^{-5/3} \nu_{\rm obs,1 \, GHz}^{1.5}$, such that for parameters characteristic of EMRI collision ejecta, 
$E_{\rm k} \sim 10^{50}$--$10^{51}$~erg, 
$M_{\rm ej} \sim 10^{-3}$--$10^{-1}~M_\odot$,
$v_{\rm w} \approx 0.05$--$0.1c$, and 
$n_{\rm ext} \sim 10^3$--$10^5~{\rm cm^{-3}}$, 
the two timescales are comparable, both in the range of
$t_{\rm pk} \sim$ few years.  

\begin{figure*}
    \centering
    \includegraphics[width=\linewidth]{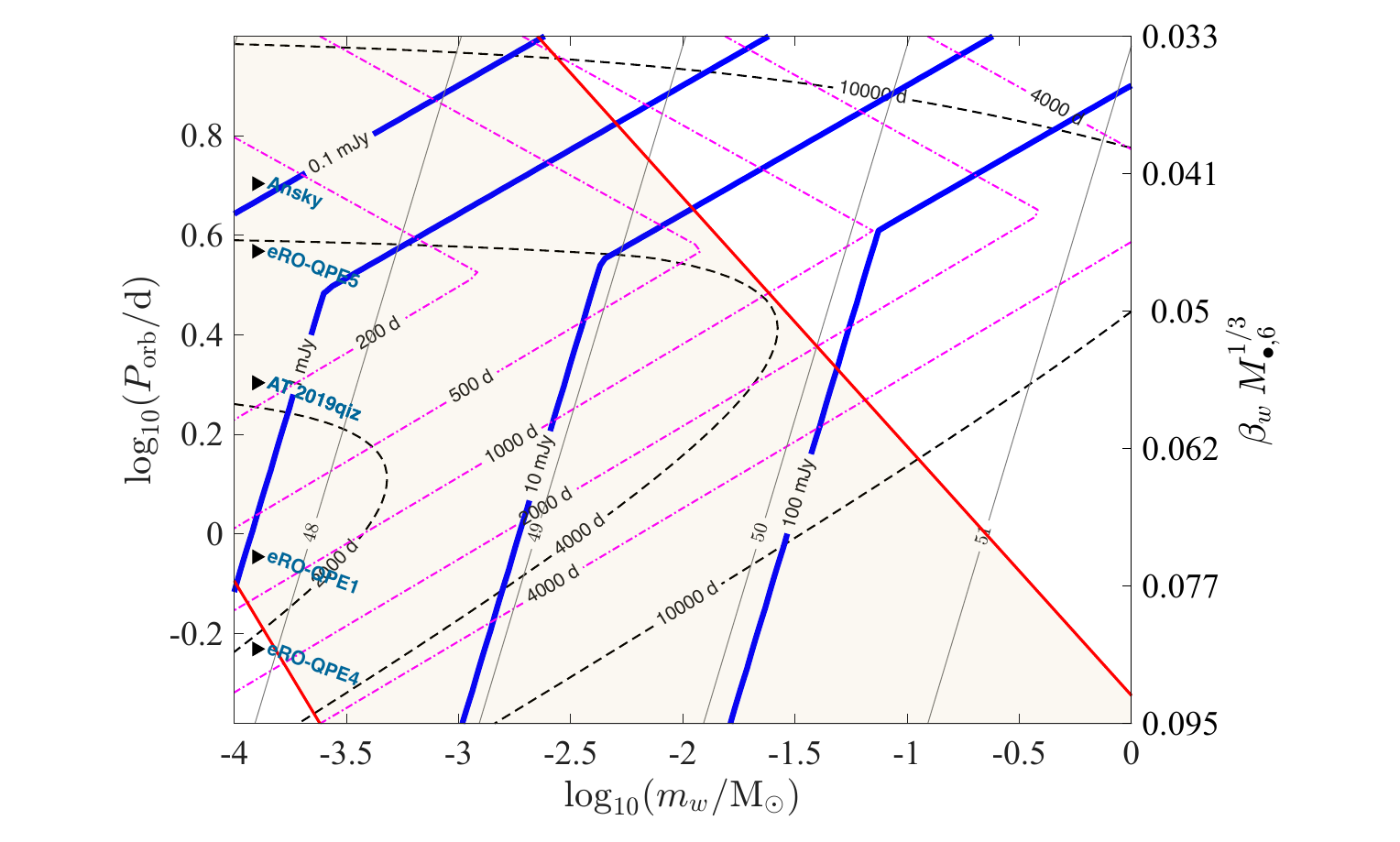}
    \caption{Contours of peak 3 GHz radio flux (for $d = 100$ Mpc; {\it solid blue}), peak timescale (measured since the outflow was launched; {\it dot-dashed magenta}), and the time of peak flux (measured since the TDE; {\it dashed black}) of radio flares from EMRI-disk collisions, as a function of the EMRI orbital period $P_{\rm orb}$ and ejecta mass $m_{\rm w}$.  Thin gray lines show the ejecta kinetic energy $\log_{10}(E_{\rm k}/\rm erg)$, while triangles on the left mark observed recurrence times of X-ray QPE sources (interpreted within the EMRI-disk collisions model).  The ejecta velocity is assumed to equal the EMRI orbital speed, as shown on the right vertical axis. The highlighted region between the two red contours brackets the expected ejecta mass $m_{\rm w}(P_{\rm orb})$, following Eq.~\eqref{eq:m_w_tot_disk} and $\eqref{eq:m_w_tot_star}$ for a range of $f_{\rm ub}^{\rm \star}\lesssim0.5$ (fixing $f_{\rm ub}^{\rm d}=0.5$, $m_\star=m_\star^{\rm tde}=1\,\rm M_\odot$). The gaseous medium surrounding the black hole is assumed to be earlier ejecta launched promptly after the TDE of mass $m_{\rm w}^{\rm e} = 0.25\, \rm M_\odot$, spherically expanding with velocity $v_{\rm w}^{\rm e} = 5\times 10^3 \, \rm km\,s^{-1}$. The delayed outflow is assumed to be launched once the spreading TDE disk reaches the EMRI orbit, with a delay $\Delta t_{\rm w} = t_{\rm v}=10^3 \, P_{\rm orb}$ after the TDE has occurred.
    Standard shock parameters are assumed: $\epsilon_e=0.1 \,, \epsilon_{\rm B}=0.01$, $p = 2.5$. }
    \label{fig:radio_obs}
\end{figure*}

Fig.~\ref{fig:radio_obs} illustrates the landscape of radio flare properties as function of $P_{\rm orb}$ and $m_{\rm w}$, under concrete assumptions regarding the EMRI/disk and the surrounding environment and the microphysical shock parameters. In particular, we show contours of the the peak 3 GHz radio flux in mJy for a typical source distance of 100 Mpc (Eqs.~\eqref{eq:Fnu1}, \eqref{eq:Fnu2}; thick blue contours), the duration of the peak measured in days since the launching time (Eq.~\eqref{eq:tpeak}; dash-dotted magenta contours), and the time to peak radio flux in days measured since the TDE (dashed black contours) as a function of the ejecta mass and EMRI orbital period.  The latter is connected in our model to the ejecta velocity based on the EMRI orbital speed, as shown along the right vertical axis.  Observed recurrence times of several X-ray QPE sources in the range of hours-days, are also shown on the left \citep{Arcodia+21,Arcodia+24a,Arcodia+25,Nicholl+24,Hernandez_Garcia_25_Ansky}.

In generating Fig.~\ref{fig:radio_obs} we assume for simplicity a uniform density medium surrounding the black hole, created by matter promptly ejected from the TDE, of mass $m_{\rm w}^{\rm e} = (1-f_{\rm d})m_\star^{\rm tde}/2$ (i.e., the bound mass that did not contribute to forming the disk), with $f_{\rm d}=0.5$, and velocity $v_{\rm w}^{\rm e} \approx 5\times 10^3 \, \rm km \, s^{-1}$. The EMRI collision outflow is launched with a delay $\Delta t_{\rm w} = t_{\rm v}(a_0)=10^3 \, P_{\rm orb}$ from the TDE, at which point the external density is ``frozen'', i.e., evaluated at the launching time, $n_{\rm ext} = 3m_{\rm w}^{\rm e}/(4\pi (v_{\rm w}^{\rm e} \Delta t_{\rm w})^3m_{\rm p}$ (and not, e.g., the actual collision time with different shells of the early ejecta). 

The highlighted region between the solid red lines indicate bracket limits on $m_{\rm w}(P_{\rm orb})$, estimated for different assumptions regarding the dominant component of the ejecta (i.e,. disk and/or ablated stellar material, Eqs.~\eqref{eq:m_w_tot_disk}, \eqref{eq:m_w_tot_star}).  Across the allowed parameter space, we predict peak flux densities $\sim 0.01-100$ mJy are achieved on timescales of typically several years.  These are largely consistent with delayed radio flares seen following TDEs \citep[][]{Cendes+24,GoodwinMummery26}, which peak on timescales of $\sim 2-10$ years at characteristic radio luminosities $\nu L_{\nu} \sim 10^{37}-10^{39}$ erg s$^{-1}$ and from which one infers (e.g., through equipartition analysis) ejecta velocities $v_{\rm w} \sim 0.02-0.15c$ and kinetic energies $\sim 10^{48}-10^{50}$ erg.

The above calculation assumes a stationary ambient medium, which is a reasonable approximation as long as the delayed outflow is much faster than the the expansion velocity of the surrounding medium. The collision energy is reduced if, for example, the delayed outflow propagates into a previously ejected early TDE outflow which has retained most of its energy. To a first approximation, one could obtain the instantaneous radio emission properties by substituting $\beta_{\rm sh}' = (v_{\rm w}-v_{\rm w}^{\rm e})/c$, and $E_{\rm k}' = E_{\rm k} (\beta'_{\rm sh}/\beta_{\rm sh})^2$. In the limit where $v_{\rm w}^{\rm e} \gg v_{\rm w}$, the delayed outflow may propagate into a cavity carved by the early ejecta. 

A more detailed treatment of outflow-outflow collisions has been discussed in \cite{Wu_2026}, who considered the radio emission from consecutive mass ejections from a thermally unstable TDE disk, undergoing repeated state transitions.  While the launching mechanism explored in their work is entirely different to ours (ejection from thermally-unstable TDE disk undergoing a state transition), the range of outflow masses and velocities they consider are similar to our findings for star-disk outflows. The agreement with observations demonstrated in their work therefore suggests that the model proposed here is likewise expected to show similar consistency with the data.

\section{Discussion and Summary}
\label{sec:implications}

\subsection{Connection to X-ray QPEs}

Repeated star-disk collisions in galactic nuclei have been invoked to explain two rare and seemingly unrelated phenomena: (1) X-ray QPEs, and (2) delayed radio flares in TDEs. Similarity in the underlying physical picture motivates an exploration of any potential links between these transients.   

Radio follow-up of known QPE sources has revealed weak emission in several cases, with tentative evidence for variability on year-long timescales \citep{Goodwin+25b}. However, no source has yet shown a clear radio rebrightening years after its discovery, as might be expected from a delayed outflow interacting with the ambient medium.

Here, we briefly highlight some of the theoretical uncertainties and questions that govern this potential connection.

\begin{itemize}
    \item Star-disk collisions are likely necessary but not sufficient for producing detectable X-ray QPEs. In particular, X-rays generated by star-disk collisions may not always be bright or hot enough to outshine the quiescent disk emission \citep{Miniutti_23a,Linial_Metzger_23}. The flare temperature depends sensitively on the collision velocity, densities of the impacted disk and/or stripped stellar debris, as well as the thermalization efficiency \citep{Linial_Metzger_23,Linial_25,Vurm+25}. 
    \cite{Linial_25} suggested that star-disk collisions may generate X-ray flares only during limited phases of the underlying TDE disk evolution, when the debris streams of matter stripped from the star possess sufficient optical depth to sustain a radiation-mediated shock.  More broadly, the criteria for ``QPE observability'' remains unclear given uncertainties in the X-ray emission mechanism.
    \item Bright radio emission requires a sufficient large-scale ambient medium surrounding the black hole, capable of decelerating the disk/star outflows. Yet, if the ambient gas is too dense, X-ray emission would also be suppressed by high scattering optical depth. In the latter case, a bright radio flare may be seen, while X-ray emission (from the disk or from star-disk collisions) is strongly attenuated. In the other extreme, a dilute ambient medium (perhaps as a result of earlier TDE ejecta clearing a void) would produce weaker radio emission.
    \item The fraction of shocked disk/star material that becomes unbound in the collision process depends on the relative inclination between the star and the disk. For low inclinations and corresponding low collision velocities $\ll v_{\rm k}$, almost none of the shocked mass will be accelerated to above the local escape speed, such that $f_{\rm ub} \to 0$. In this scenario, the launched outflow may be much weaker than nominally assumed throughout this paper. Remarkably, if QPEs arise from impacts of a stellar-mass black hole EMRI (rather than a star, e.g., \citealt{Tagawa&Haiman23,Franchini_23,Lam+25}), low inclination trajectories are favored for producing the observed QPE luminosities (since low relative velocity imply a larger Bondi-Hoyle radius, increasing the amount of mass impacted with every disk passage). This scenario would suggest that the presence of X-ray QPEs disfavors the launching of powerful outflows, and vice-versa.
\end{itemize}

Despite these uncertainties, QPE observations already provide constraints on outflows from disk–orbit interactions. For example, the QPE source ZTF19acnskyy \citep[``Ansky'',][]{Hernandez_Garcia_25_Ansky} exhibits blueshifted absorption features with velocities $v/c \sim 0.1$ \citep{Chakraborty+25b}, interpreted as requiring ejecta masses of order $\delta m\sim 10^{-3} \, \rm M_\odot$ per flare. Within the framework of this paper, such events imply outflow rate $\dot{m}_{\rm w} \sim f_{\rm ub} \times 0.03 \, {\rm M_\odot \, yr^{-1}} \, (P_{\rm orb}/10 \, \rm d)^{-1}$, and total kinetic energy $\sim 10^{51} \, \rm erg$ assuming an evolution time of $\sim 1 \, \rm yr$ (comparable with the observed orbital evolution timescale, $|P/\dot{P}|$; \citealt{Chakraborty+26}).

In general, the outflow properties of QPE sources can be roughly constrained from the flare energy and recurrence time. If a fraction $\varepsilon_{\rm rad}^{\rm qpe} \lesssim 1$ of the kinetic energy of the shocked gas is radiated as soft X-rays with energy $E_{\rm fl} = 10^{47}E_{\rm fl,47}$ erg per flare, the required shocked mass obeys:
\begin{multline}
    \delta m_{\rm sh} \approx \frac{2E_{\rm fl}}{v_{\rm k}^2 \varepsilon_{\rm rad}^{\rm qpe}} \approx \\
    2\times 10^{-5} \, {\rm M_\odot}\, E_{\rm fl,47} P_{\rm orb,24}^{2/3} M_{\bullet,6}^{-2/3}/\varepsilon_{\rm rad}^{\rm qpe} \,,
\end{multline}
implying an average outflow rate
\begin{multline}
    \dot{m}_{\rm w} \approx \frac{f_{\rm ub} \delta m_{\rm sh}}{P_{\rm orb}/2} \approx \\
    0.017 \, {\rm M_\odot \, yr^{-1}} \,\frac{f_{\rm ub}}{\varepsilon_{\rm rad}^{\rm qpe}}  E_{\rm fl,47}  P_{\rm orb,24}^{-1/3} M_{\bullet,6}^{-2/3} \,,
\end{multline}
such that the total ejecta mass over the viscous time $t_{\rm v}$ is approximately
\begin{equation} \label{eq:m_w_tot_qpe}
    \dot{m}_{\rm w} t_{\rm v} \approx 
    0.04 \, {\rm M_\odot}
    \frac{f_{\rm ub}}{\varepsilon_{\rm rad}^{\rm qpe}}  E_{\rm fl,47}  P_{\rm orb,24}^{2/3} M_{\bullet,6}^{-2/3} \left(\frac{t_{\rm v}}{10^3 P_{\rm orb}}\right) \,,
\end{equation}
and a total kinetic energy
\begin{equation}
    E_{\rm k} \approx 10^{50}\,{\rm erg}\, \, \frac{f_{\rm ub}}{\varepsilon_{\rm rad}^{\rm qpe}} E_{\rm fl,47} (t_{\rm v}/10^3 P_{\rm orb}) \,.
\end{equation}

\subsubsection{Predictions for AT 2019qiz}

AT 2019qiz provides a compelling case study linking TDEs, QPEs, and delayed outflows. Early observations revealed a prompt outflow of velocity $3\times10^3-10^4 \, \rm km \,s^{-1}$, likely responsible for the initial radio flare \citep{Nicholl+20}. Several years later, repeating soft X-ray flares (QPEs) were discovered with a recurrence time $P_{\rm QPE} \approx 48 \, \rm hr$ \citep{Nicholl+24}, indicating the onset of QPEs with a delay $\Delta t^{\rm qpe} \approx  4 \, \rm yr$ after the TDE. 

Interpreting the QPE flares from AT 2019qiz as repeated star-disk collisions requires an EMRI orbital period $P_{\rm orb}=P_{\rm QPE}$ (or $P_{\rm orb}=2 P_{\rm QPE}$, if two flares are observed per orbit). Assuming the collisions launch an outflow of velocity comparable to the Keplerian orbital speed, $v_{\rm w} \sim v_{\rm k} \approx 0.07 \, c \, M_{\bullet,6.3}^{1/3} P_{\rm orb,48}^{-1/3}$, the collision of this otuflow with the ambient medium (e.g., the prompt TDE outflow) may power a bright delayed, radio flare.

The predicted energy and mass of the late outflow can be estimated from the QPE properties. Based on the average energy radiated per QPE flare in AT 2019iz, $E_{\rm fl}^{\rm qiz} \approx 5.8\times 10^{47} \, \rm erg$, the inferred ejecta mass associated with each flare obeys $\delta m^{\rm qiz} \approx 2(\varepsilon_{\rm rad}^{\rm qpe})^{-1} E_{\rm fl}^{\rm qiz}/(v_{\rm k}^{\rm qiz})^2 \approx (\varepsilon_{\rm rad}^{\rm qpe})^{-1}\,10^{-4} \, {\rm M_\odot}$, where $\varepsilon_{\rm rad}^{\rm qpe} \lesssim 1$ is the radiative efficiency.  If the TDE disk formed well inside the EMRI's orbit, then we have an upper limit on the disk's viscous spreading time, $t_{\rm v} \lesssim \Delta t_{\rm qpe}$. Substituting into Eq.~\ref{eq:m_w_tot_qpe}, we predict an outflow mass $m_{\rm w} \lesssim f_{\rm ub} 0.2 \, {\rm M_\odot}$ and kinetic energy $E_{\rm k} \approx f_{\rm ub} \, 10^{50.5} \, \rm erg$, where $f_{\rm ub}$ is again the unbound ejecta fraction.

Although AT 2019qiz was among the QPE sources discussed in \cite{Goodwin+25b}, their reported radio detection occurred in 2019-2020, shortly after the optical discovery. This radio emission therefore likely associated with early TDE outflow \citep{Nicholl+20}, rather than delayed ejections from EMRI-disk collisions nominally associated with the X-ray QPEs, first observed in 2024. Additional radio data for this source was recently published by \citet[their Table 7]{Alexander+26}, indicating a possible rise in radio flux at around $t\approx 1200 \, \rm d$, roughly when the X-ray QPEs were discovered from this source. If star-disk collisions did not commence much earlier (although see \citealt{Alush&Stone25}), their delayed outflow were also launched around this epoch. We therefore predict that the radio emission from AT 2019qiz will rebrighten as the EMRI-induced outflow sweeps through the ambient medium over a timescale of years. 

If the surrounding density is low (e.g., because the early TDE outflow cleared a cavity), the delayed radio emission will peak on the deceleration time, $t_{\rm dec} \approx 9 \, {\rm yr} \, n_2^{-1/3}$ (measured from the onset of star-disk collisions, i.e., approximately 2024). We predict flux density $F_{\nu,\rm 19qiz}^{\rm max} \approx 7 \, {\rm mJy} \, \epsilon_{e,-1}^{1.5} \epsilon_{\rm B,-2}^{0.88} n_2^{0.87} f_{\rm ub}$ at $\nu_{\rm obs}=1\,\rm GHz$. Conversely, the absence of radio rebrightening would place tight constraints on QPE-related outflows, their gaseous environments, and the efficiency of mass ejection.

\subsection{Rates}

The fraction of TDEs which exhibit delayed radio flares depends on how frequently TDEs occur in the presence of a star or compact object on a short-period orbit.

A common channel to form stellar EMRIs is via binary disruption near the SMBH \citep{Hills_1988,Linial_Sari_2023}. In the empty loss-cone regime, this occurs at a rate comparable to the TDE rate, up to a factor given by the fraction of stars paired as a tight binaries. Depending on whether such EMRIs survive repeated disk crossings, this implies that $\sim 1\%-100\%$ of TDEs could host a nearby EMRI (\citealt{Linial_Metzger_23}).

Observationally, roughly a dozen QPE sources have been discovered to date, including in the aftermath of the optically selected TDEs AT 2019qiz and AT 2022upj \citep{Nicholl+24,Chakraborty+25} and the X-ray TDE J2344 \citep{Baldini+26}. The QPE-TDE association therefore constrains the prevalence of EMRIs interacting with TDE disks. Current estimates suggest that about $\sim 10\%$ of TDEs are expected to host X-ray QPEs, in rough agreement with the above theoretical considerations \citep{Chakraborty+25}. This is lower than the $\sim 40\%$ fraction of TDEs which exhibit delayed radio flares \citep{Cendes+24}. This discrepancy may indicate that only a subset of delayed radio flares arise from EMRI–disk interactions, that not all EMRI systems produce observable QPEs, or that a larger population of TDEs host QPEs which remain undetected due to limited X-ray followup resources.

\subsection{Radio Emission following TDEs}

\subsubsection{WTP14adeqka}
\cite{Golay+25} reported a late-time radio flare seen in the mid-IR TDE WTP14adeqka, several years after the IR peak. This is the first source where the delayed radio flare is spatially resolved using the VLBA, providing powerful constraints on the size and evolution of the synchrotron emitting region, independent of equipartition assumptions. The inferred outflow properties are $E_{\rm k} \approx 10^{50.7} \, \rm erg$, $\beta_{\rm w} \approx 0.05$, and launching time $\Delta t_{\rm w} \approx 500-10^3 \, \rm d$. The shocked mass, $m_{\rm w} \gtrsim 0.1 \, {\rm M_\odot}$, is substantially higher than in previously analyzed radio flares \citep[e.g.,][, Fig. 6]{Cendes+24}. However, energy estimates from equipartition analysis represent a lower limit on the real energy/mass of the outflow, and actual value may be closer to those inferred for WTP14adeqka. The mass swept up by the outflow is comparable to the initial ejecta mass at the deceleration time, $t_{\rm dec}$, which is not well-constrained by the current data. Future VLBA observations that better measure the source's expansion velocity and deceleration history could improve constraints on the initial ejecta mass and velocity.

Star-disk interaction may produce such an outflow in the presence of an EMRI with orbital period $P_{\rm orb} \approx 10-20 \, {\rm d}$ interacting with the TDE disk (considering the inferred SMBH mass of $10^{6.8} \, \rm M_\odot$, \citealt{Masterson_24}). A relatively fast evolving disk, $t_{\rm v}/P_{\rm orb} \sim \Delta t_{\rm w}/P_{\rm orb} \approx 10^2$, is required to match the observed launch time (unless the initial TDE disk size was substantially larger, perhaps due to a more massive TDE progentior star). As WTP14adeqka was discovered as an IR TDE, the system is likely engulfed by a high dust content, which would suppress X-ray emission, precluding the possibility of observing X-ray QPEs in this source, even if star-disk collisions are ongoing. 

\subsection{Survivability of TDE disks and QPE sources} \label{sec:TDE_QPE_Survivability}

The multi-band quiescent emission in several QPE sources is well-described by TDE-like compact accretion disks (e.g., \citealt{Wevers_2025_ero2,Guolo_2025_GSN069_joint,Nicholl+24,Chakraborty+25}). In both GSN 069 and eRO-QPE2, the disk's outer radius lies well outside an orbit whose period matches the observed QPE recurrence time \citep{Guolo_2025_GSN069_joint,Wevers_2025_ero2}, i.e., $r_{\rm d} \gg a_0$. If initially, $r_{0} < a_0$, the system must have evolved through an intermediate phase where $r_{\rm d} \approx a_0$, lasting approximately $t_{\rm v}(a_0)$. During this phase, when the local surface density $\Sigmad(a_0)$ is maximal, and the EMRI sustains the highest amount of ablation. The survival of the EMRI at late times, when $r_{\rm d} \gg a_0$ thus depends on the ratio $t_{\rm abl}/t_{\rm v}$ (Eq.~\eqref{eq:t_abl_over_t_v}).

An additional regime which we did not directly address earlier in the text is when $a_0 < r_{\rm d,0}$ -- i.e., the EMRI is interior to the TDE circularization radius. Star-disk collisions commence shortly after disk formation, possibly launching an outflow contemporaneously with other early TDE outflows. $\Sigmad(a_0)$ is then maximal during the initial disk spreading time, $t_{\rm v,0} \approx r_{\rm d,0}^2/\nu_0$. The survivability of the EMRI can then be expressed as
\begin{equation} \label{eq:t_abl_over_t_v_inner}
    \frac{t_{\rm abl}}{t_{\rm v}(r_{\rm d,0})} \approx 0.3 \, \frac{P_{\rm orb,8}^{5/3}} {f_{\rm d}  \eta_{0.03} (t_{\rm v}/10^3 P) m_1^{0.4} (m_{\star,1}^{\rm tde})^{0.77}} \,,
\end{equation}
where we approximated $\Sigmad^{\rm max} \approx f_{\rm d} m_\star^{\rm tde}/(\pi (2\rtidal^{\rm tde})^2)$ as the initial disk surface density. Short period EMRIs with $a_0 < 2\rtidal^{\rm tde}$ are thus more susceptible to be significantly ablated following a TDE.

In the case of eRO-QPE2, a stellar-EMRI is permitted only for $m_1 \lesssim 0.2 \, \rm M_\odot$ given its short orbital period of $P_{\rm orb}^{\rm eRO2}\simeq 5 \, \rm hr$ (e.g., \citealt{Linial_25}). Setting $t_{\rm abl}/t_{\rm v} > 1$ in Eq.~\ref{eq:t_abl_over_t_v_inner} thus implies $t_{\rm v}/P_{\rm orb} < 260 \, f_{\rm d}^{-1} (m_{\star,1}^{\rm tde})^{0.77}$, at $r=r_{\rm d,0}$. Namely, the disk must evolve sufficiently fast to avoid destruction of the EMRI. A similar survivability argument in the context of GSN 069 requires $t_{\rm v}/P \lesssim 10^3$ for fiducial parameters and $P_{\rm orb}^{\rm GSN} \approx 18 \, \rm hr$. Interesingly, time-dependent modeling of GSN 069's quiescent multi-band emission has yielded a stronger limit, $t_{\rm v}/P_{\rm orb}(r_{\rm d}) \lesssim 230$ \citep{Guolo_2025_GSN069_time_dependent}, compatible with the EMRI survivability argument.

Another possible outcome of star-disk interaction is the depletion of the accretion disk by collision-induced mass ejection. This regime, illustrated in Fig.~\ref{fig:OutcomesRegimes} has interesting observational implications. As the outflow is being launched over the disk depletion time, $t_{\rm dep}$, the disk X-ray/UV/optical emission would subside. Thus, some TDE sources that show delayed radio flares might also show a fading UV/optical plateau, on a timescale comparable to the outflow launching timescale.

\subsection{Summary}
\label{sec:conclusions}

Late-time radio flares in TDEs, appearing years after the optical peak, are now frequently observed. The events are generally attributed to delayed outflows with velocities $v_{\rm w} \approx 0.01-0.1 \, c$, and masses $m_{\rm w} \approx 10^{-3}-1 \, \rm M_\odot$, launched hundreds to thousands of days after the disruption. However, the inferred mass of these outflows can rival or even exceed the mass remaining in the accretion disk at such late times, placing disk-fed outflow models under potential tension (see Fig.~\ref{fig:OutflowMassRadioTDEs} and discussion in Sec. \ref{sec:intro}). This tension is further exacerbated by the continued presence of disk-dominated optical/UV emission after the outflow is launched, which limits the fraction of disk mass that can be ejected. 

Here we have proposed a mechanism in which delayed outflows are powered by interactions between the TDE disk and a stellar EMRI, which provides an additional mass reservoir that alleviates the disk mass budget constraint. In this picture, star-disk collisions do not commence until the TDE disk expands sufficiently to intercept the EMRI's orbit, thereby delaying the launch of an outflow. The outflow's velocity is set by the star-disk collision speed, of the order of the EMRI's Keplerian velocity, while the ejected mass originates from ablation of the star and the intercepted disk material.

Our proposed mechanism predicts a close connection between delayed radio flares and TDEs that have occurred in the presence of an EMRI. In particular, we draw a possible connection between delayed radio flares and X-ray QPEs that follow a TDE. We discuss the possibility of a joint detection of both QPEs and a delayed radio flare, with specific emphasis on sources like AT 2019qiz, AT 2022upj and J2344, where QPEs have been recently identified in the aftermath of optical/X-ray TDEs. 

\begin{acknowledgments}
We would like to thank Matt Nicholl, Yvette Cendes, Kate Alexander, Andrew Mummery, Adelle Goodwin, Philippe Yao, Eliot Quataert, Indrek Vurm and Riccardo Arcodia for useful discussions. Support for this work was provided by NASA through the NASA Hubble Fellowship grant \#HST-HF2-51581.001-A awarded by the Space Telescope Science Institute, which is operated by the Association of Universities for Research in Astronomy, Inc., for NASA, under contract NAS5-26555. This research benefited from interactions at workshops funded by the Gordon and Betty Moore Foundation through grant GBMF5076. This research was also partly funded through Simons Investigator grant 827103. BDM acknowledges support from NASA (grant 80NSSC24K0934), the National Science Foundation (grant AST-2406637), and the Simons Foundation (grant 727700). The Flatiron Institute is supported by the Simons Foundation. AMB acknowledges support by NASA grant 80NSSC24K1229, NSF grant AST-2408199, and Simons Foundation grant 446228. This work was also facilitated by Multimessenger Plasma Physics Center (MPPC) grant PHY-2206609.

\end{acknowledgments}

\appendix

\section{Glossary} \label{sec:Glossary}
\begin{deluxetable*}{lll}
\tablecaption{Glossary of symbols\label{tab:Glossary}.}
\tablehead{
\colhead{Symbol} &
\colhead{Description} &
\colhead{Equation}
}
\startdata
$\MBH$
& SMBH mass
& - \\
$m_\star \,, R_\star$
& EMRI mass, radius
& - \\
$m_\star^{\rm tde} \,, R_\star^{\rm tde}$
& TDE progenitor mass, radius
& - \\
$\rtidal$
& Tidal radius
& - \\
$P_{\rm orb} \,, a_0 \,, v_{\rm k}$
& EMRI orbital period, semi-major axis, orbital speed
& \ref{eq:v_k_from_P} \\
$m_{\rm d} \,, r_{\rm d} \,, \Sigmad \,, \nu \,, \dot{m}_{\rm d} \,, t_{\rm v}$
& Disk mass, radial size, surface density, viscosity, accretion rate, viscous time
& \ref{eq:m_disk_at_a0} \\
$f_{\rm d}$
& Fraction of bound debris forming a disk
& \ref{eq:m_disk_at_a0} \\
$\delta m_{\rm d} \,, \delta m_\star$
& Shocked disk mass, ablated stellar mass, per collision
& \ref{eq:delta_m_disk}, \ref{eq:delta_m_abl} \\
$\eta$
& Ablation efficiency
& \ref{eq:delta_m_abl} \\
$f_{\rm ub}^{\rm d} \,, f_{\rm ub}^\star$
& Unbound fraction of $\delta m_{\rm d}$, $\delta m_\star$
& - \\
$\xi_{\rm w}$
& Ejection efficiency coefficient
& \ref{eq:xi_w_gen} \\
$k$
& Disk mass conservation parameter
& \ref{eq:sink} \\
$m_{\rm w} \,, \dot{m}_{\rm w} \,, v_{\rm w}$
& Outflow mass, ejection rate and typical velocity
& \ref{eq:mdot_gen} \\
$\Delta t_{\rm w}$
& Delay since TDE until outflow is launched
& - \\
$t_{\rm dec}$, $t_{\rm SSA}$
& Deceleration timescale of launched outflow, synchrotron self-absorption time
& \ref{eq:t_dec}, \ref{eq:t_SSA} \\
\enddata
\end{deluxetable*}

\bibliography{main}
\bibliographystyle{aasjournal}

\end{document}